\documentclass[aps,prd,nofootinbib,superscriptaddress]{revtex4-1} 
\usepackage{graphics}
\usepackage[english]{babel}
\usepackage{amsmath} 
\usepackage{verbatim} 
\usepackage{mathrsfs}
\usepackage{amsfonts}
\usepackage[latin1]{inputenc}
\usepackage{hyperref}

\newcommand{\pyRI}{\hat{\pi}^{R,I}}

\newcommand{\fluc}[1]{(\Delta #1)^2}      
 
\newcommand{\cre}{\hat{a}^{\dagger}}    

\newcommand{\ann}{\hat{a}}            

\newcommand{\tc}{\eta_{\vec{k}}^c}
\newcommand{\beq}{\begin{equation}}
\newcommand{\eeq}{\end{equation}}
\newcommand{\nk}{\vec{k}}
\newcommand{\dphi}{\delta \phi}
\newcommand{\x}{\vec{x}}

\newcommand{\bra}{\langle}
\newcommand{\ket}{\rangle}
\newcommand{\bea}{\begin{eqnarray}}
\newcommand{\eea}{\end{eqnarray}}

\newcommand{\almsix}{\overline{a_{l_1 m_1} a_{l_2 m_2} a_{l_3 m_3} a_{l_1 m_4}^\star a_{l_2 m_5}^\star a_{l_3 m_6}^\star}}

\begin{document}

\author{Gabriel Le\'on}
\email{gleon@df.uba.ar}
\affiliation{Departamento de F\'isica, Facultad de Ciencias Exactas y Naturales, Universidad de Buenos Aires and IFIBA, CONICET, Ciudad Universitaria, Buenos Aires 1428, Argentina}

\author{Daniel Sudarsky}
\email{sudarsky@nucleares.unam.mx}
\affiliation{Instituto de Ciencias Nucleares, Universidad Nacional Aut\'onoma de M\'exico, M\'exico D.F. 04510, M\'exico}

\title{Origin of structure: Primordial Bispectrum without non-Gaussianities}

\begin{abstract}
The primordial bispectrum has been considered in the past decade as a powerful probe of the physical processes taking place  in the early Universe. Within the inflationary paradigm, the properties of the bispectrum are one  of the keys  that serves to discriminate among competing scenarios concerning the details  of the origin of cosmological perturbations.  However,  all of the scenarios, based on the conventional approach to  the so-called  ``quantum-to-classical transition'' during  inflation, lack the ability to point out the precise  physical mechanism responsible for generating the inhomogeneity and anisotropy of our Universe starting  from and exactly homogeneous and isotropic vacuum state associated with the early inflationary regime. In past works, we have shown that the proposals  involving a spontaneous  dynamical reduction of the quantum state provide plausible explanations for the birth of said  primordial inhomogeneities and anisotropies. In the present letter, we show that, when considering 
single-field 
slow-roll inflation  within the   context of   such  proposals, the 
 expected characteristics of the  bispectrum turn out to be quite different from those  found in the traditional approach. In particular, the statistical features corresponding to the primordial perturbations, which are normally associated with the bispectrum, are treated  here  in a novel way leading to  rather  different   conclusions.
\end{abstract}



\maketitle
\section{Introduction}\label{Intro}

Recent advances in observational cosmology  are allowing a detailed   testing of  various models regarding the early Universe. In particular, the inflationary paradigm, considered one of the most promising models for the primordial Universe, is considered  as  providing an explanation for  the origin of cosmological perturbations \cite{mukhanov1990,hawking1982,starobinsky1982,guth1982,halliwell1985,mukhanov1992}. Indeed,  recent observational data (e.g. WMAP \cite{wmap9}, SDSS \cite{sdss9}, \emph{Planck} \cite{planck}), is in    rather  broad of agreement with the theoretical predictions offered  by  the  inflationary  paradigm. According to  this theory, the idea behind the generation of the primordial inhomogeneities, ``the seeds of galaxies,'' is also rather image-evoking: The perturbations start as quantum fluctuations of the inflaton field, as the Universe experiments a phase of accelerated expansion, the physical wavelength associated with the perturbations is stretched out reaching  cosmological 
scales.
 At this point one is 
invited to treat the quantum fluctuations as classical density perturbations. Then,  the   argument goes,  at later cosmological epochs, these perturbations continue evolving into the cosmic structure responsible for galaxy formation, stars, planets   and eventually  life and  human beings.

However, as  has  been discussed  at length  in  previous works,  the  complete theory must not only  allow  one to  find  expressions  that are in agreement with observations, but also, be able to provide an  explanation of  the   precise  physical   mechanism behind  its predictions. 
As originally discussed in \cite{sudarsky2006}, there is a  conceptual difficulty  in the standard explanation for the birth of cosmic structure provided by inflation, this is, from a highly\footnote{At  the  level of  one part in  $e^{N}$  where $N$  is the number of  e-folds of inflation.} homogeneous and isotropic state that characterizes both: the initial  state of the  so-called   quantum  perturbations and   the classical background  inflaton  and space-time, the Universe ends in a state with ``actual" inhomogeneities and anisotropies. In other words, if we consider quantum mechanics as a fundamental theory applicable in particular  to the  Universe as  a whole; then  any classical descriptions must be regarded as imprecise characterizations of  complicated quantum mechanical states. The Universe that we observe today is clearly well described by an inhomogeneous and anisotropic classical 
state; therefore, such description must be considered as an  imperfect description of an equally inhomogeneous and anisotropic quantum state. Consequently, if we want to consider the  inflationary  account  as  providing the physical  mechanism  for the generation of  the seeds of structure,   such account must contain an explanation for why the quantum state that describes our actual Universe does not possess the same symmetries  as  the early quantum state of the Universe, which happened to be perfectly symmetric (the symmetry being the homogeneity and isotropy). Since there is nothing in the dynamical evolution (as given by the standard inflationary approach) of the quantum state that can break those symmetries, then we are left with an incomplete theory. In fact, this and other shortcomings have been recognized by others in the literature \cite{Padmanabhan,Mukhanov,Weinberg,Lyth}.

The detailed discussion of the conceptual problems associated with the inflationary paradigm, and its possible solutions following the standard rules provided by Quantum Mechanics (e.g., the decoherence program, many-worlds interpretation and the consistent histories approach),  have been presented by some of us and by others in  \cite{susana2013,pearle2007,sudarsky2006,sudarsky2009}. We will not reproduce those arguments here and invite the interested reader to consult those references;  the  above  paragraph  is  meant  only to provide the reader to  a small indication of the  kinds of issues  that the  detailed  analysis  of such questions  involves\footnote{ In fact,   even if one  wants  to adopt,  say, the many-worlds  interpretational posture,  the issue  at  hand can be rephrased by  asking  questions  about the precise  nature of the  quantum state that can be taken  as representing our  specific   branch  of the many  worlds. One can also  focus on the issue  of  characterizing in a precise  
mathematical way the  quantities that encode  the  so-called stochastic  aspects, which are often only  vaguely referred  to.  We  will see this   in a very concrete  way  in the  the following.}

The idea that has been presented in previous works \cite{sudarsky2006,sudarsky2006b,sudarsky2007,sudarsky2009,sudarsky2011}, as a possibility to deal with the aforementioned problem, involves supplementing the standard inflationary model with  an  hypothesis  concerning the modification of quantum  theory including  a  spontaneous  dynamical  reduction   of the quantum state  (sometimes referred  as  the self-induced  collapse of the wave function) and consider it as an actual physical process; taking place  independently of observers or measuring devices. 
Regarding the situation at hand, the basic scheme is the following: A few $e$-folds after inflation has started, the Universe finds itself in an homogeneous and isotropic quantum state, then during the inflationary regime a quantum collapse of the wave function is triggered (by  novel physics possibly related with quantum gravitational effects),  breaking in the process the unitary evolution of quantum mechanics and also, in   general, breaking the symmetries of the original state. The post-collapse state continues to evolve leading to  one state  that is  not isotropic  or homogeneous and,  moreover, it   is  susceptible  to  an approximate  classical  characterization describing  a  Universe, which includes the inhomogeneities and anisotropies, that will unfold, due to standard physical processes, into  what we observe today.

The hypothesis regarding the self-induced collapse is not a new idea and there has been a considerably amount of research along this lines:  The continuous spontaneous localization (CSL)  model \cite{pearle1989}, representing  a   continuous  version of the Ghirardi-Rimini-Weber (GRW) model \cite{ghirardi1985}, and the ideas of Penrose \cite{penrose} and Di\'osi \cite{diosi} regarding gravity as the main agent responsible for the collapse, are among the main programs proposed to describe the physical mechanism of a self-induced collapse of the wave function. For more recent examples  see Refs. \cite{weinberg2011,bassi2003}.
In fact, the implications of applying the CSL model to the inflationary scenario have been studied in Refs. \cite{pedro,jmartin} and \cite{tpsingh} leading to interesting results   constraining  the parameters of the model in terms of   the parameters  characterizing  the inflationary  model.

On the other hand,  the statistical analyses  in  Refs. \cite{adolfo,susana2012} show  how the predictions of the simple collapse schemes,  used  in  previous  works,  can be  confronted with recent data from the Cosmic Microwave Background (CMB), including the 7-yr release of WMAP \cite{wmap7} and the matter power spectrum measured using LRGs by the Sloan Digital Sky Survey \cite{sdssb}. In fact, results from those analyses indicate that  while  several schemes or ``recipes'' for the collapse are compatible with the observational data,   others  are not, allowing to establish constraints on the free parameters of the schemes. Those  works serve to underscore the point   that,   in addressing the conceptual issues of the inflationary paradigm, one  is not   only dealing   with ``philosophical issues,'' but that   these  have  impact in the theoretical predictions. While,   at the same time,  the conclusions  drawn  can  lead  to  important insights,   as well as  a better  understanding  of the nature  of  
its predictions  and  a novel way to  consider the relation  of  these predictions  with   the observations.

In this  work,  we will be primarily concerned with the characteristics of   the ``primordial bispectrum.''  In the inflationary paradigm, the primordial bispectrum has been regarded as  one of the indicators  characterizing   any primordial ``non-Gaussianities.''  By non-Gaussianity, one    generically refers to any  deviations in the observed fluctuations from the random field of linear, Gaussian, curvature perturbation. 
It is commonly believed that the study of non-Gaussianities will play a leading role in furthering our understanding of the physics of the very early Universe that created the primordial seeds for large-scale structures \cite{komatsu2009}. As a matter of fact, the shape and amplitude of the bispectrum is  used  nowadays to discriminate among various inflationary models. The amplitude of non-Gaussianity is 
characterized in terms of a dimensionless  parameter $f_{\textrm{NL}}$ (defined in Sec. \ref{diferencias}). Distinct models of inflation predict different values for  $f_{\textrm{NL}}$. The amount of  non-Gaussianity from simple inflationary models, which are based on a single slow-roll scalar field, is expected  to be very small \cite{salopek1990,salopek1991,falk1992,gangui1993,acquaviva2002,maldacena2002}; however, a very large class of more general models with, e.g., multiple scalar fields, special features in inflaton potential, non-canonical kinetic terms, deviations from Bunch-Davies vacuum, among others (see Ref. \cite{bartolo2004} for a review and references therein),  are thought to be able  to generate substantially higher amounts of non-Gaussianity.\footnote{For more details and derivations regarding non-Gaussianity, we refer the reader to the comprehensive review by Komatsu \cite{komatsu2003}, Bartolo et al. \cite{bartolo2004} and others \cite{Yadav,Liguori,Komatsu2001}.}

In this article, we will consider the simplest slow-roll inflation model, within  the  approach   involving  the hypothesis of ``the collapse of the wave function.''  We  will  show  that despite   the fact  that the  scheme  leads  to linear Gaussian  curvature perturbations,   our proposal induce correlations that,  under   the standard  approach,  would  end up  being  identified  as   non-Gaussianities. This is due to the fact that   the   same field  modes, which   undergo independent collapses,  contribute  to the various   coefficients of   spherical  harmonic  decomposition  of the  CMB  anisotropy pattern. Our analysis is  made possible because, in our approach,  we must use   explicit    variables  characterizing  the random aspects  that   determine   the primordial bispectrum.\footnote{At this point,   and in order to avoid any  misconceptions,   we  must  warn the reader  familiar with the subject    that we are not   talking  here  about the quantum expectation value of the three point function,
  but   that we  are  focusing on  the quantity that  the  observers  actually measure, and   which, in   our view,  differs  from the former.} In other words, assuming a Gaussian metric perturbation, our proposal predicts a non-vanishing value for the   observational bispectrum. The underlying reason for this,  seemingly  self-contradictory result, is that the statistical aspects leading to the estimate  for  the observable   bispectrum, within the collapse proposal, differ dramatically  from the  corresponding  expectations  of  the  conventional approach. Therefore, although the 
primordial bispectrum offer us an 
observational tool to understand the physical processes of the early Universe, the statistical aspects  that are involved   in the  comparison of theory and observations,  must, in  our view,   be addressed   in  a different manner. 

 In addition to the primordial bispectrum, we will analyze a new quantity (previously introduced in Ref. \cite{susana2013}) that offers two advantages with respect of the former. 
First, the observational and theoretical notions are clearly separated; this contrasts with the common treatment for obtaining an observational value for the amplitude and shape of the bispectrum. Second, it serves to illustrate that maintaining clear distinctions between each of the averages involved (quantum averages, ensemble averages, time-space averages and orientation averages)  yields a  different  set of predictions for the observational quantities.
As  we  will  find,  in the traditional picture the   expected value for the new quantity vanishes exactly, while working in the approach  based  on  collapse framework, one is lead  to  expect a  non-vanishing value  for this  quantity.

The rest of the paper is organized as follows: In Section \ref{review}, we start by reviewing the ideas and technical aspects of our proposal, in particular we focus on how to implement 
the collapse of the wave function during the inflationary Universe. Then, in Section \ref{seccolapsobispec}, we derive an analytic estimate of the  expected value  of the  observed 
primordial bispectrum within our approach. Afterwards, in Section \ref{diferencias}, we discuss the main differences between the collapse proposal and the standard approach regarding 
the estimates  of  the  primordial bispectrum and its statical aspects.  Section \ref{nuevacantidad} contains a detailed analysis for a novel observational quantity that allows us to 
differentiate between the two approaches, both, at the quantitative and qualitative level. In Section \ref{comparisons}, we present a discussion regarding the comparisons between our theoretical prediction and the observational data. Finally, in Section \ref{concl}, we end with a discussion of our conclusions. We have also included 
two Appendixes \ref{calculos}, \ref{apendiceb} that present   the details of the  more lengthy calculations appearing  in Secs. \ref{seccolapsobispec} and \ref{nuevacantidad}. 

Regarding conventions and notation, we will be using a $(-,+,+,+)$ signature for the space-time metric. The prime over the functions  $f'$ denotes derivatives with respect to the conformal time $\eta$. We will use units where $c=\hbar=1$ but will keep the gravitational constant $G$.

\section{Brief review of the collapse proposal within the inflationary Universe}\label{review}

In this section we will present a brief review of the collapse proposal which has been exposed in great detail in previous works \cite{sudarsky2006,gabriel2010,adolfo,adolfo2010}. The main purpose is to present  the central ideas behind the proposal in order to make the presentation as self-contained as possible. 
The full  self-consistent formalism  was developed in \cite{alberto}; however, we will not use  here such full fledged formal treatment. 
This is because, as can be seen there, the problem becomes extremely cumbersome even in the treatment of a single mode of the inflationary field. Thus, 
when studying the CMB bispectrum, the task  would  quickly become a practical impossibility. Instead, we will focus on the   simpler  pragmatical approach first proposed in \cite{sudarsky2006}.   More details   of this approach  can be   found  in  \cite{gabriel2010}. In  fact,  we  believe that  the   analysis  we  will preform here,     would have analogous   counterparts  in other  much  more developed  approaches involving dynamical collapse   theories,  such  as the    GRW \cite{ghirardi1985}  or CSL \cite{pearle1989}  proposals,    and   even  in approaches that rely on applying Bohmian Mechanics  to the cosmological   problem  \cite{valentini,pintoneto}.  For examples using CSL  in this context  see \cite{jmartin, tpsingh, pedro}.

Before proceeding with the technical aspects, we wish to discuss  the way the gravitational sector and  the matter fields are   treated in our  approach. As we have  not yet at our disposal  a fully workable  and satisfactory theory of quantum gravity, we will rely on the ``semi-classical gravity'' approach which  of course  we  take  only as   an effective  setting  rather than  something that can  be  consider as  a  fundamental theory. The inflationary period is assumed to start at energy scales smaller than the planck mass ($\sim 10^{-2} M_P$), thus, the semi-classical approach is a suitable approximation  for something that, in principle,  ought to  be treated   in   a precise  fashion  within a quantum gravity theory. The semi-classical framework is characterized by Einstein semi-classical equations $G_{ab} = 8 \pi G \bra \hat{T}_{ab} \ket$, which  allow to relate the quantum degrees of freedom of the matter fields with the classical description of gravity in terms of the metric.\footnote{During the 
collapse, the 
semi-classical approximation will not remain  
  100\% valid; this is because the quantum collapse would induce sudden changes or ``state jumps" to the initial quantum state, thus the divergence $\nabla_a \bra \hat{T}^{ab} \ket \neq 0$ while $\nabla_a G^{ab} = 0$. However, as we will be only interested in the states \emph{before} and \emph{after} the collapse, this breakdown of the semi-classical approximation would not be important for our present work.} The use of such semi-classical picture has two main conceptual advantages:

First, the description and treatment of the metric is always ``classical." As a consequence there is no issue with the ``quantum-to-classical transition" in the sense that one needs to justify going from ``metric operators" (e.g. $\hat \Psi$) to classical metric variables (such as $\Psi$).  The fact that the space-time remains classical is particularly important in  the context  of models   involving dynamical reduction of the wave function,  as   such    ``collapse or reduction" is  regarded   as {\it a  physical process taking place in time}  and,  therefore,   it is  clear  that  a  setting allowing  consideration of    full space-time  notions  is  preferred  over, say,  the  ``timeless"  settings usually  encountered  in canonical  approaches to quantum gravity (for  some basic references on ``the problem of time  on quantum gravity''  see Ref. \cite{TIME}).

Second, it allows to present a transparent picture of how the inhomogeneities and anisotropies are born from the quantum collapse: the initial state of the Universe (i.e. the one characterized by a few $e$-folds after inflation has started) is described by the homogeneous and isotropic Bunch-Davies vacuum, and the equally homogeneous and isotropic classical Friedmann-Robertson-Walker space-time. Then, at a later stage, the quantum state of the matter fields reaches a phase whereby the corresponding state for the gravitational degrees of freedom are forbidden, and a quantum collapse of the matter field wave function is triggered by some unknown physical mechanism. In this manner, the state resulting from the collapse needs not to share the same symmetries as the initial state. After the collapse, the gravitational degrees of 
freedom are assumed to be, once more, accurately described by Einstein semi-classical equation. However, as $\bra \hat{T}_{ab} \ket$ for the new state needs not to have the symmetries of the pre-collapse state, we are 
led to a geometry that generically will no longer be homogeneous and isotropic.

We proceed now to introduce the details of the  simplest  collapse proposal. The starting point in our approach is the same as the standard slow-roll inflationary model; this is one writes the action of a scalar field (the inflaton) minimally coupled to gravity, 

\beq\label{accioncolapso}
S[\phi,g_{ab}] = \int d^4x \sqrt{-g} \bigg( \frac{1}{16 \pi G} R[g] - \frac{1}{2} \nabla_a \phi \nabla_b \phi g^{ab} - V[\phi] \bigg).
\eeq

Einstein's field equations $G_{ab} = 8 \pi G T_{ab}$ are derived from \eqref{accioncolapso}, with $T^a_b$ given by

\beq
T^a_b = g^{ac} \partial_c \phi \partial_b \phi + \delta^a_b \left( -\frac{1}{2} g^{cd}\partial_c \phi \partial_d \phi - V[\phi] \right).
\eeq

The next step is to split the metric and the scalar field into a background plus perturbations $g_{ab} = g_{ab}^{(0)} + \delta g_{ab}$, $\phi = \phi_0 + \dphi$. The background is represented by a spatially flat FRW space-time with line element $ds^2 = a(\eta)[-d\eta^2 + \delta_{ij}dx^i dx^j]$ and the homogeneous part of the scalar field $\phi_0 (\eta)$. We will choose $a=1$ at the present cosmological time; while we assume that the inflationary period ends at a conformal time $\eta_\star \simeq -10^{-22}$ Mpc.

The scale factor corresponding to the inflationary era is $a(\eta) \simeq -1/(H \eta)$ with $H$ the Hubble factor defined as $H \equiv \partial_t a/a$, thus $H \simeq$ const. During inflation $H$ is related to the inflaton potential as $H^2 \simeq (8 \pi G /3) V$. The scalar field $\phi_0(\eta)$ is in the slow-roll regime, which means that $\phi_0' \simeq -(a^3/3a') \partial_\phi V$. The slow-roll parameter defined by $\epsilon \equiv \frac{1}{2} M_P^2 (\partial_\phi V/V)^2$ is considered to be $\epsilon \ll 1$;  $M_P$ is the reduced \emph{Planck} mass defined as $M_P^2 \equiv 1/(8 \pi G)$.


Next,  we focus on the perturbations. Ignoring the vector and tensorial perturbations, and working  in the conformal Newtonian gauge, the perturbed space-time is represented by

\beq
ds^2 = a(\eta)^2 [-(1+2\Phi) d\eta^2 + (1- 2\Psi)\delta_{ij}dx^idx^j],
\eeq
with $\Phi$ and $\Psi$ functions of the space-time coordinates $\eta,x^i$. Einstein's equations to first order in the perturbations lead to $\Phi = \Psi$ and

\beq\label{master}
\nabla^2 \Psi = 4\pi G \phi_0' \dphi' =- \sqrt{\frac{\epsilon}{2}} \frac{aH}{ M_P} \dphi',
\eeq
where in the second equality we used Friedmann's equations and the definition of the slow-roll parameter.

Next, we consider the quantization of the theory. As  mentioned  above,  we will  work  within the  collapse-modified  semi-classical gravity 
setting (for  a detailed  discussion of the self-consistent formalism see Ref. \cite{alberto}). In particular, we will quantize the fluctuation of the inflaton field $\dphi (\x,\eta)$,  but not the  metric perturbations. For simplicity, we will work with the rescaled field variable $y=a\dphi$. One then proceeds to expand the action \eqref{accioncolapso} up to second order in the rescaled variable (i.e. up to second order in the scalar field fluctuations)

\beq\label{acciony}
\delta S^{(2)}= \int d^4x \delta \mathcal{L}^{(2)} = \int d^4x \frac{1}{2} \left[ y'^2 - (\nabla y)^2 + \left(\frac{a'}{a} \right)^2 y^2 - 2 \left(\frac{a'}{a} \right) y y' \right].
\eeq
The canonical momentum conjugated to $y$ is $\pi \equiv \partial \delta \mathcal{L}^{(2)}/\partial y' = y'-(a'/a)y=a\dphi'$.

In order to avoid distracting infrared divergences, we set the problem in a finite box of side $L$. At the end of the calculations we can take the continuum limit by taking $L \to \infty$. The field and momentum operators are decomposed in plane waves

\beq
\hat{y}(\eta,\x) = \frac{1}{L^3} \sum_{\nk} \hat{y}_{\nk} (\eta) e^{i \nk \cdot \x}  \qquad \hat{\pi}(\eta,\x) = \frac{1}{L^3} \sum_{\nk} \hat{\pi}_{\nk} (\eta) e^{i \nk \cdot \x},
\eeq
where the sum is over the wave vectors $\vec k$ satisfying $k_i L=
2\pi n_i$ for $i=1,2,3$ with $n_i$ integer and $\hat y_{\nk} (\eta) \equiv y_k(\eta) \ann_{\nk} + y_k^*(\eta)
\cre_{-\nk}$ and  $\hat \pi_{\nk} (\eta) \equiv g_k(\eta) \ann_{\nk} + g_{k}^*(\eta)
\cre_{-\nk}$. The function $y_k(\eta)$ satisfies the equation:

\beq\label{ykmov}
y''_k(\eta) + \left(k^2 - \frac{a''}{a} \right) y_k(\eta)=0.
\eeq
To complete the quantization, we have to specify the mode  solutions of \eqref{ykmov}.  The canonical commutation relations between $\hat y$ and $\hat \pi$, will  give $[\hat{a}_{\nk},\hat{a}^\dag_{\nk'}] = L^3 \delta_{\nk,\nk'}$, when $y_k(\eta)$ is  chosen  to satisfy $y_k g_k^* - y_k^* g_k = i$ for all $k$ at some time $\eta$.

The remainder of  the  choice of $y_k(\eta)$ corresponds to the so-called Bunch-Davies  (BD) vacuum, which is characterized by

\beq
y_k(\eta) = \frac{1}{\sqrt{2k}} \left( 1- \frac{i}{\eta k} \right) e^{-ik\eta}, \qquad g_k(\eta) = -i \sqrt{\frac{k}{2}} e^{-ik\eta}.
\eeq
There is certainly  some   arbitrariness  in  selection of a  natural  vacuum state,     but it seems  clear that any such  natural choice would be spatially a homogeneous and isotropic state. The  BD vacuum  certainly is a homogeneous and isotropic state as can be seen by evaluating directly the action of a translation or rotation operator on the state.

From $G_{ab} = 8 \pi G \bra \hat{T}_{ab} \ket$ and \eqref{master} it follows  that

\beq\label{master2}
\Psi_{\nk} (\eta) = \sqrt{\frac{\epsilon}{2}} \frac{H}{M_P k^2} \bra \hat{\pi}_{\nk} (\eta) \ket.
\eeq
It is clear from Eq. \eqref{master2} that if the state of the field is the vacuum state, the metric perturbations vanish, and, thus the space-time is homogeneous and isotropic.

The self-induced collapse model is based on considering that the collapse operates very similar to a  kind of self-induced ``measurement" (evidently, there is no external observer or detector involved). In considering  the operators  used to characterize the post-collapse  states,  it seems  natural therefore  
to focus  on Hermitian operators,  which in ordinary quantum mechanics are the ones susceptible of direct measurement. 
We thus  separate $\hat y_{\nk} (\eta)$ and $\hat \pi_{\nk} (\eta)$ 
into  their  ``real and imaginary parts" $\hat y_{\nk} (\eta)=\hat y_{\nk}{}^R (\eta) +i \hat y_{\nk}{}^I (\eta)$ and $\hat \pi_{\nk} (\eta) =\hat \pi_{\nk}{}^R (\eta) +i \hat \pi_{\nk}{}^I (\eta)$ . The point is that the operators $\hat y_{\nk}^{R, I} (\eta)$ and $\hat \pi_{\nk}^{R, I} (\eta)$ are  hermitian. Thus,\footnote{$\mathcal{R}[z]$ 
denotes the real part of 
$z \in \mathbb{C}$} $\hat{y}_{\nk}^{R,I} (\eta) = \sqrt{2} \mathcal{R}[y_k(\eta) \hat{a}_{\nk}^{R,I}]$, $\hat{\pi}_{\nk}^{R,I} (\eta) = \sqrt{2} \mathcal{R}[g_k(\eta) \hat{a}_{\nk}^{R,I}]$,
where $\hat{a}_{\nk}^R \equiv (\hat{a}_{\nk} + \hat{a}_{-\nk})/\sqrt{2}$, $\hat{a}_{\nk}^I \equiv -i (\hat{a}_{\nk} - \hat{a}_{-\nk})/\sqrt{2}$.
 The commutation relations for the $\hat{a}_{\nk}^{R,I}$ are non-standard

\beq\label{creanRI}
[\hat{a}_{\nk}^R,\hat{a}_{\nk'}^{R \dag}] = L^3 (\delta_{\nk,\nk'} + \delta_{\nk,-\nk'}), \quad [\hat{a}_{\nk}^I,\hat{a}_{\nk'}^{I \dag}] = L^3 (\delta_{\nk,\nk'} - \delta_{\nk,-\nk'}),
\eeq
with all other commutators vanishing.

Following the a  line of   thought  described  above, we  assume that the collapse is somehow analogous to an imprecise measurement\footnote{An imprecise measurement of an observable is 
one in which one does not end with an exact eigenstate  of that observable, but  rather with a state that is  only peaked around the eigenvalue. Thus, we could consider measuring a  
certain particle's position and momentum so as to end up with a state that is a wave packet with both position and momentum defined to a limited extent and, which certainly, does not  entail a conflict with Heisenberg's uncertainty bound.}
of the
operators $\hat y_{\nk}
^{R, I}
(\eta)$ and $\hat \pi_{\nk}
^{R, I}
(\eta)$.
The rules according to which the collapse  is assumed to  happen are guided by simplicity and naturalness.

In particular,   as  we  are  taking the view that a collapse  effect on  a state is analogous  to some  sort of  approximate measurement, we  will postulate   that  after the  collapse, the expectation values of 
the field and momentum operators  in each mode  will  be related to the uncertainties  of the  initial  state. For the purpose of this  work  we will  work with a particular collapse scheme called  the  
\emph{Newtonian} collapse scheme which is given by\footnote{In previous works, we have analyzed other collapse schemes such as the \emph{independent} scheme and the 
\emph{Wigner} scheme. See Refs. \cite{sudarsky2006,adolfo,susana2012} for detailed analyses of the collapse schemes.}

\begin{equation}
\langle {\hat{y}_{\nk}^{R,I}
(\eta^c_k)} \rangle_\Theta = 0  
\label{momentito}
\end{equation}

\begin{equation}
\langle {\hat{\pi}_{\nk}{}^{R,I}
(\eta^c_k)}\rangle_\Theta
= x^{R,I}_{\nk}\sqrt{\fluc{\pyRI_{\nk}}
_0} ,
=  x^{R,I}_{\nk}|g_k(\eta^c_k)|\sqrt{L^3/2},
\label{momentito1}
\end{equation}
where $\tc$ represents the \emph{time of collapse} for each mode. In the vacuum state,  $\hat{\pi}_{\nk}$ is distributed according to a Gaussian wave function centered at $0$ with spread  
$\fluc{\hat{\pi}_{\nk}}_0$.  The motivation for choosing such scheme is two-folded. First, the calculations performed for this scheme are relatively easier to handle. Second, in Eq. 
\eqref{master2} the variable that is directly related with the Newtonian Potential $\Psi$ is the expectation value of $\hat \pi$; therefore, it seems natural to consider that the variable 
affected at the time of collapse is $\bra  \hat \pi_{\nk} (\tc) \ket$  while $\bra y_{\nk} (\tc) \ket = 0$.

The  random variables $x_{\nk}^{R,I}$   represent  values  selected randomly 
from  a  Gaussian distribution    with  unit dispersion.
At this point, we must emphasize that our Universe 
corresponds to a \textbf{single realization} of these random variables, and thus  each of these quantities
$ x^{R}_{\nk}$, $ x^{I}_{\nk}$ has a  single specific value. It is  clear that  even though  we  will not do that here,  one could   also investigate  how  the statistics  of   $x_{\nk}^{R},x_{\nk}^{I}$ might be affected by the physical 
process of the collapse\footnote{In a recent work \cite{sigma} (motivated by the findings of \cite{alberto}),  we explored a correlation between the random variables  of any mode  with those of their higher harmonics (something reminiscent of the  so-called  parametric   resonances  found  in  quantum optics in materials  with  nonlinear   response  functions \cite{Parametric-resonance}). As mentioned in \cite{sigma}, we found that this effect leads to a departure from the standard prediction afflicting only the first multipoles of the angular power spectrum; in fact, something 
resembling to our findings has been observed in newest results obtained from \emph{Planck} satellite \cite{planckcmb} and will be the subject of future research.} (e.g. see Ref. \cite{sigma}).
 The statistics  of  these  quantities can  be studied   using as  a tool an imaginary ensemble of ``possible Universes,'' 
 but we  should   in principle distinguish   those   from  the statistics of  such  quantities for the 
particular  Universe  we inhabit; we will discuss these and other aspects in the next sections.

The next step is to find an expression for the evolution of the expectation values of the field operators at  all  times. 
  This  can be  done in various  ways   but the simplest invokes  using Ehrenfest's  theorem  
to obtain  the expectation values of the field operators  at  any later time    in terms of the expectation values at the time of 
collapse. The result is:

\beq\label{expecyeta}
\bra \hat{y}_{\nk}^{R,I} (\eta) \ket_\Theta = \bigg[ \frac{\cos (k\eta - z_k)}{k} \bigg( \frac{1}{k\eta} - \frac{1}{z_k} \bigg) + \frac{\sin (k\eta - z_k)}{k} \bigg(\frac{1}{k\eta z_k} + 1 \bigg) \bigg]  \bra 
\hat{\pi}_{\nk}^{R,I} (\tc) \ket_\Theta,
\eeq

\beq\label{expecpieta}
\bra \hat{\pi}_{\nk}^{R,I} (\eta) \ket_\Theta  = \bigg( \cos (k\eta - z_k) +\frac{\sin (k\eta - z_k)}{z_k} \bigg) \bra \hat{\pi}_{\nk}^{R,I} (\tc) \ket_\Theta,  
\eeq
with $z_k \equiv k\tc$. This calculation is explicitly done in Refs. \cite{sudarsky2006,adolfo2010}.

Finally, using \eqref{master2},  \eqref{momentito1}  and \eqref{expecpieta} we find and expression for the Newtonian potential in terms of the random variables and the time of collapse 

\beq\label{masterrandom}
\Psi_{\nk} (\eta) =     \frac{ \sqrt{\epsilon}  H}{ M_P} \left( \frac{L}{2k} \right)^{3/2}   \left( \cos (k\eta - z_k)+ \frac{\sin (k\eta - z_k)}{z_k} \right) X_{\nk}, 
\eeq
where $X_{\nk} \equiv x_{\nk}^R + i x_{\nk}^I$. This last expression is the main result of the present section. It relates the Newtonian potential during inflation to the parameters 
describing the collapse (i.e. the random variables and the time of collapse). It is worth noting that all the quantities occurring  in \eqref{masterrandom} are  all classical quantities and   no quantum operators  appear in the  expression. This is an  important   difference between our approach and the standard treatment  of perturbations  during    inflation. That is,  in 
the latter approach, the Newtonian potential is strictly 
a quantum operator and then   one needs to  invoke various   kinds of arguments that are  often  vague  and  do  not  lead to clear connections  with the quantities found in the observations; in particular,    there is  often  an  appeal to  quantum  randomness  that is,    however, left   completely unspecified. 
Thus,   the standard   approach   suffers  from the lack  of opportunity  for  clear  characterization of the stochastic aspects  of the   situation  (as  well  as   from  other  conceptual  deficiencies  that  have  been have discussed  in \cite{sudarsky2009}).
In our approach, we  will not   rely on  arguments  involving horizon-crossing of the modes, decoherence or many worlds interpretation of 
quantum mechanics, 
to justify the transition from a quantum object $\hat \Psi$ to a classical stochastic field $\Psi$, which result  in a rather vague  connection  of the  mathematical  expressions used  and  the   objects 
that emerge  from  observations.  One of the  advantages of the approach  we favor is that, as a result of the  collapse  postulate, such connection  become   transparent and  specific:   we  have the variables  $X_{\nk}$  characterizing,   once  and  for all, every stochasticity   we  will  need  to deal with.

As is well known, the Newtonian potential is closely  related with the  the temperature anisotropies  whose origins  can be traced back (in the specific gauge)  with the extra  red/blue shift  photons  suffered  when emerging from the local potential wells/hills. As  the values of the two random 
variables associated to each mode, $x_{\nk}^{R}$ and $x_{\nk}^{I}$, are fixed for our Universe, it follows  from  expression \eqref{masterrandom} that  these values determine   the  value  of the Newtonian potential Fourier components  corresponding to  our  Universe, which in turn fix the value of the   observed temperature anisotropies.  The statistic nature of  the prescribed  distribution of the random variable $X_{\nk} \equiv x_{\nk}^{R}+ i x_{\nk}^{I} $ gets transfered to the Newtonian potential $\Psi_{\nk}$; if the random variable is Gaussian, then $\Psi_{\nk}$ is also Gaussian. It is   clear that we cannot give a definite prediction for the values that  these random variables take in our Universe, given the  intrinsic randomness of  the collapse. 
 However,  as  we will show next,   the fact that we have  a large number  of modes  $\vec k$   contributing to each of the observed  quantities,  will allow  us to perform  a statistical  analysis and obtain theoretical estimates for the observational  quantities.

\section{Characterizing the primordial CMB bispectrum}\label{seccolapsobispec}

In the first subsection, we will study the connection between the Newtonian potential at the end of inflation and the observational quantities obtained from the temperature anisotropies 
of the CMB; in particular, we will show how the temperature fluctuations are related with the collapse parameters.  In the second subsection, we will provide the connection between the 
parameters characterizing the collapse and the primordial \textit{bispectrum}.


\subsection{Observational quantities}\label{angularspectrum}

The observational quantity of interest corresponds to the temperature fluctuations of the CMB observed today on the celestial two-sphere. The temperature anisotropies are expanded using the spherical harmonics $ \frac{\delta T}{T_0} (\theta,\varphi) = \sum_{l,m} a_{lm} Y_{lm} (\theta,\varphi)$, which means that the coefficients $a_{lm}$ can be expressed as

\beq\label{alm0}
a_{lm} = \int \frac{\delta T}{T_0} (\theta,\varphi) Y_{lm}^\star (\theta,\varphi) d\Omega,
 \eeq
here $\theta$ and $\varphi$ are the coordinates on the celestial two-sphere, with $Y_{lm}(\theta,\varphi)$ the spherical harmonics ($l=0,1,2...$ and $-l \leq m \leq l$), and $T_0 \simeq 2.725$ K the temperature average.

The different multipole numbers $l$ correspond to different angular scales; low $l$ to large scales and high $l$ to small scales. At large angular scales ($l \leq 20$), the Sachs-Wolfe effect is the predominant source to the temperature fluctuations in the CMB. That effect relates the anisotropies in the temperature observed today on the celestial two-sphere to the inhomogeneities in the last scattering surface,

\beq
\frac{\delta T}{T_0} (\theta,\varphi) \simeq \frac{1}{3} \Psi^{\text{matt}} (\eta_D, \x_D),
\eeq
where $\x_D= R_D (\sin \theta \sin \varphi, \sin \theta \cos \varphi, \cos \theta)$, with $R_D$ the radius of the last scattering surface and $\eta_D$ is the conformal time of decoupling ($R_D \simeq 4000$ Mpc, $\eta_D \simeq 100$ Mpc). The Newtonian potential can be expanded in Fourier modes, $\Psi^{\text{matt}} (\eta_D, \x_D) = \sum_{\nk} \Psi_{\nk}^{\text{matt}} (\eta_D) e^{i \nk \cdot \x_D} / L^3$ . Furthermore, using that $e^{i \nk \cdot \x_D} = 4 \pi  \sum_{lm} i^l j_l (kR_D) Y_{lm} (\theta,\varphi) Y_{lm}^\star (\hat k )$,  expression \eqref{alm0} can be rewritten as

\beq\label{alm1}
a_{lm} = \frac{4 \pi i^l}{3L^3} \sum_{\nk} j_l (kR_D) Y_{lm}^\star(\hat k) \Psi_{\nk}^{\text{matt}} (\eta_D),
\eeq
with $j_l (kR_D)$ the spherical Bessel function of order $l$.

The Newtonian potential $\Psi^{\text{matt}}$ appearing in \eqref{alm1} is evaluated at the time of decoupling which corresponds to the matter dominated cosmological epoch. Traditionally, the relation between $\Psi^{\text{matt}}$ and the Newtonian potential at the end of inflation is made by making use of the so-called transfer functions $T(k)$; the transfer functions 
contain all relevant physics from the end of inflation to the latter matter dominated epoch, which includes among others the acoustic oscillations of the plasma. Thus, $\Psi_{\nk}^{\text{matt}} (\eta_D) = T(k) \Psi_{\nk}$,  where $\Psi_k$ corresponds to the Newtonian potential during inflation and, since one is interested in the modes with scales of observational interest,  $\Psi_{\nk}$ correspond to the limit $-k\eta \to 0$ of $\Psi_{\nk} (\eta)$ (or, as commonly referred, its scale should be ``well outside the horizon'' during inflation).  This is, the coefficients $a_{lm}$ are rewritten as

\beq\label{alm2}
a_{lm} = \frac{4 \pi i^l}{3L^3} \sum_{\nk} j_l (kR_D) Y_{lm}^\star(\hat k) T(k) \Psi_{\nk}.
\eeq

At this point the traditional approach would proceed to calculate averages and higher-correlation functions of the coefficients $a_{lm}$. Nevertheless, within our model we can make a further step, by substituting Eq. \eqref{masterrandom} (and taking the limit $-k\eta \to 0$), which gives an explicit expression for $\Psi_{\nk}$ in terms of the parameters of the collapse, in Eq. \eqref{alm2} one obtains\footnote{Note that we have multiplied by a factor of $3/(5 \epsilon)$ the $\Psi_{\nk}$ we obtained during inflation, Eq. \eqref{masterrandom}. This is because, while $\Psi_{\nk} (\eta)$ is constant for modes $-k\eta \ll 1$ during any cosmological epoch, its behavior changes substantially during a change in the equation of state for the dominant type of matter in the Universe. In particular, during the change from inflation to radiation epochs, $\Psi$ is amplified by a factor of approximately $1/\epsilon$. For a detailed discussion regarding the amplitude within the collapse framework see Ref. \cite{
gabriel2010}. 
 }

\beq\label{almrandom}
a_{lm} = \frac {  i^l}{ L^{3/2}} \sum_{\nk} \frac{ g(z_k) j_l (kR_D) Y_{lm}^{\star} (\hat k)}{k^{3/2}} T(k) X_{\nk},
\eeq
where

\beq\label{gz}
g(z_k) \equiv  \frac{\pi H}{5 M_P} \sqrt{\frac{2}{\epsilon}} \bigg( \cos z_k   - \frac{\sin z_k}{z_k} \bigg).
\eeq

Equation \eqref{almrandom} allow us to appreciate one of the advantages of the collapse proposal: the coefficient $a_{lm}$, which is directly associated with the observational quantities (i.e. the temperature fluctuations), is in turn related to the random variables characterizing the collapse. In other words, the statistical features of the coefficients $a_{lm}$ can 
be discussed   in terms of  the statistics of the random variables $X_{\nk}$. 
We note   that  there is no analog expression of Eq. \eqref{almrandom} in the 
standard approach. As a matter of fact, if we follow the conventional way of identifying quantum expectation values with classical quantities, the prediction given by the standard inflationary paradigm would be $\bra 0 | \hat \Psi_k |0\ket = 0 = \Psi_k$; thus, we  would be lead, by Eq. \eqref{alm2}, to conclude that
 $a_{lm}=0$; this is, the theoretical prediction for the temperature fluctuations would be exactly zero in an evident contradiction\footnote{Several kinds of  arguments would normally  be invoked  at this point  in  defense of the standard treatments.  For a  detailed discussion of their merits and  shortcomings  see Ref. \cite{sudarsky2009}.} with the observations (see Ref. \cite{susana2013}).

One  key   aspect that  in  our  treatment  differs,  from those followed  in the   standard   approaches,  is  the manner in which  the  results  from the formalism  are   connected  to observations.  This  is    most clearly  exhibited  by   our result regarding the quantity  $a_{lm} $   in Eq. \eqref {almrandom}.  Despite the fact that we have in principle  a close    expression for   the quantity of interest,  we  cannot use Eq.  \eqref {almrandom}  to make  a definite prediction  because   the  expression involves the  numbers   $ X_{\nk}$  that correspond, as  we  indicated  before,  to   a    random    choice  ``made  by  nature''   in the context of the collapse process. 
 The  way one  make  predictions  is by regarding  the  sum  appearing   in Eq.   \eqref {almrandom}   as   representing a   kind of  two-dimensional  random walk,  i.e the sum   of complex numbers   depending on    random  choices  (characterized by the $ X_{\nk}$).  As  is  well known, for a random  walk,  one cannot   predict the final displacement (which  would correspond to the    complex  quantity $a_{lm} $),    but  one  might estimate  the  most likely value of the  magnitude  of such  displacement.  Thus,  we focus  precisely  on the  most likely  value of $|a_{lm}|$,  which we  denote  by   $|a_{lm}|_{\text{M.L.}}$.    In order to compute  that  quantity,  we   make  use  of a   fiducial  (imaginary)  ensemble of  realizations of the random  walk  and  compute the    ensemble   average value over    of the  total displacement.  Thus  we   identify:
 
 \beq\label{ML}
|a_{lm}|_{\text{M.L.}} = \overline{|a_{lm}|}.
\eeq

The overline appearing denotes average over the    fiducial   ensemble  of realizations,    which  would correspond  to  an  imaginary ``ensemble of universes.''
    
The   estimate is  done  now  in the standard  way  in  which  one  deals   with  such  random  walks:
 
 \beq\label{ML2}
|a_{lm}|^2_{\text{M.L.}} = \overline{ |a_{lm}|^2}= \frac {1}{ L^{3}} \sum_{\nk, {\nk}'} \frac{ { g(z_k) j_l (kR_D) Y_{lm}^{\star} (\hat k) T(k)}
{ g(z_{k'}) j_l (k'R_D) Y_{lm} (\hat k') T(k')} } {k^{3/2} k'^{3/2}}  \overline{X_{\nk} X^\star_{{\nk}'}},
\eeq
which  upon  using the   normalized gaussian  assumption for  fiduciary  ensemble   ($ \overline{X_{\nk} X^\star_{{\nk}'}} = 2 \delta_{\nk, {\nk}'} $),  leads to

   \beq\label{ML3}
|a_{lm}|^2_{\text{M.L.}} = \frac {2}{ L^{3}} \sum_{\nk,} {k^{-3}}  j_l (kR_D)^2 | Y_{lm} (\hat k)|^2 T(k)^2 g(z_k)^2 . 
   \eeq
   
 Finally, we can remove the   fiducial  box of  side $L$   and  pass to the continuum

   \beq\label{ML4}
|a_{lm}|^2_{\text{M.L.}} 
= \int \frac{d^3k}{ 4\pi^3 k^3}    j_l (kR_D)^2 | Y_{lm} (\hat k)|^2 T(k)^2 g(z_k)^2  .
    \eeq
  At   this  point, one   could  focus   on   the   quantity  that   is  most  often studied  in this context, namely
     
   \beq\label{ML5}
 C_l  \equiv \frac{1}{2l+1}  \sum_m  |a_{lm}|^2
    \eeq
 for which  we  would  have the estimate
      \beq\label{ML6}
 {C_l}^{\text{M.L.}}  \equiv \frac{1}{2l+1}  \sum_m  |a_{lm}|_{\text{M.L.}}^2 = \frac{1}{4\pi^3} \int_0^\infty  \frac{dk}{ k } j_l (kR_D)^2 T(k)^2  {g (z_k)^2}.
    \eeq
   where in the last  step,   we  used the fact that   the     most  likely value   estimate  in Eq.  \eqref{ML4}  is  independent of  $m$. Furhtermore, if we consider the time of collapse as $\tc \propto k^{-1}$, i.e. $z_k = z$ independent of $k$ and take $T(k)=1$, which is a valid approximation for $l \ll 20$,  we recover an exact scale-invariant spectrum, this is, 
   
   \beq\label{CL}
   l(l+1) C_l^{\text{M.L.}} = \frac{g(z)^2}{(2\pi)^3} = \frac{H^2}{10^2 \pi M_P^2 \epsilon } \left( \cos z - \frac{\sin z}{z} \right)^2 \equiv A
   \eeq
   The quantity $A$ is fixed by the observational data to be $A \simeq 10^{-10}$. The fact that $\tc \propto k^{-1}$ is also motivated by the results in  previous works \cite{sudarsky2006,gabriel2010,adolfo,adolfo2010,alberto2012}

   If we would like to recover the full angular spectrum, one should then use expression \eqref{ML6} including the transfer functions, which can be obtained using numerical codes, and assume a particular form for the time of collapse $\tc$ in terms of $k$. This type of studies have been done, and the results can be consulted in Ref. \cite{susana2012}.

   The expression above is the  theoretical estimate  to be compared  with the observational    data, and  as   should be clear from  the discussion,  the fact that we  have to rely on   most likely  values,  for what  are    in  effect    the  mathematical equivalent of random walks,  leads us to     expect  that there  should  be  a   general and rough  agreement    between our estimates and  observations (assuming the theory is  correct). However, we  do not really   expect    a    detailed  and precise  match    simply  due to the  intrinsic  randomness   involved. In the  standard  approach,    similar considerations involving the  randomness of the fluctuations  and the   uncertainties   tied to  stochasticity,   and  with the limited  region of the Universe  one  is  observing,   also leads to 
  people in the community  to   expect   small  differences in predictions  and observations. Nevertheless, in general, such   discussions    are based on heuristic   arguments;  therefore,  are  limited   both in scope and precession.  The  essential   difficulty is that, in the standard  analysis,  the    precise  stochastic  elements  are not  identified and have no  mathematical   representation in the  formalism.   We  believe  that,  in our  approach,   the stochastic  elements    are  clearly identifiable    (i.e.  the  $ X_{\nk}$). This  represents  a great  advantage providing  us,  for instance, with  an explicit expression for the  quantity $ a_{lm}  $,  such as  in  Eq. \eqref{almrandom},  and, thus,  allowing us to study in great detail   the  precise  nature of   higher  order statistical   estimates    as  we  will  do  in the following.

\subsection{The primordial bispectrum}\label{colapsobispec}

The usual path to look for  non-trivial statistical features (e.g. possible non-Gaussianities) in the CMB is to study the \emph{bispectrum}, which is  considered  as related with the three-point function of the temperature anisotropies in harmonic space. The CMB angular bispectrum is defined as

\beq\label{bispec0}
B^{l_1 l_2 l_3}_{m_1 m_2 m_3} \equiv \overline{a_{1_1 m_1} a_{1_2 m_2} a_{1_3 m_3}}.
\eeq

The overline appearing in \eqref{bispec0} denotes average over an ensemble of universes, but in practice it is taken as an average over orientations in our own Universe; the relation 
between the two types of averages is not  clear and  direct (this fact has been discussed in   great detail  in Ref. \cite{susana2013}). In the following, we will show how our approach helps to clarify certain 
issues  that   emerge when dealing with the statistical aspects of the spectrum and  when comparing  theoretical   estimates  and    observations.


Given the definition of the CMB bispectrum and considering a rotational invariant sky, one finds in the literature \cite{komatsu2003,komatsu2009} another object called the ``angle-averaged bispectrum'' defined by

\beq\label{avg}
B_{l_1 l_2 l_3} \equiv \sum_{m_i}  \left( \begin{array}{lcr}
      l_1& l_2 & l_3  \\
     m_1 & m_2 & m_3
    \end{array}
    \right) B^{l_1 l_2 l_3}_{m_1 m_2 m_3} = \sum_{m_i}  \left( \begin{array}{lcr}
      l_1& l_2 & l_3  \\
     m_1 & m_2 & m_3
    \end{array}
    \right) \overline{a_{1_1 m_1} a_{1_2 m_2} a_{1_3 m_3}}.
\eeq
The object $ \left( \begin{array}{lcr}
      l_1& l_2 & l_3  \\
     m_1 & m_2 & m_3
    \end{array}
    \right) $ is called the Wigner 3-$j$ symbol (see Ref. \cite{Komatsu2001} for more details and properties for these functions) and is non-vanishing for the values of  $l,m$  satisfying  the following conditions:

\begin{enumerate}
 \item $m_1 + m_2 + m_3 = 0$.
 
 \item $ l_1+l_2 + l_3$ is an integer, (or an even integer if $m_1=m_2=m_3=0$). 
 
 \item  $|l_i-l_j| \leq l_k \leq l_i +l_j$ for all permutations of indices.
 
\end{enumerate}
These conditions are called ``the triangle conditions'' as $l_1,l_2,l_3$ must correspond to the sides of a triangle. As a matter of fact, in the standard approach, one intents to estimate $B_{l_1 l_2 l_3}$ from the observational data (e.g. see  Sec. 3.1  of Ref. \cite{planckng}) by testing different configurations for such ``triangles.''

Motivated by the fact that within the collapse proposal we can obtain a direct relation between the coefficients $a_{lm}$ and the random variables characterizing the collapse [Eq. \eqref{almrandom}] we will be  focussing  on the  expression for   the  ``observational"   bispectrum:  

\beq\label{bispecorig}
\mathcal{B}^{\text{obs}}_{l_1 l_2 l_3} \equiv \sum_{m_i}  \left( \begin{array}{lcr}
      l_1& l_2 & l_3  \\
     m_1 & m_2 & m_3
    \end{array}
    \right) a_{1_1 m_1} a_{1_2 m_2} a_{1_3 m_3} .
\eeq
 Note that this object also contains the Wigner 3-$j$ symbol, therefore,  unless $l_1,l_2,l_3$ satisfy the three previous conditions, the observational  bispectrum will vanish. The difference between $B_{l_1 l_2 l_3}$ and $\mathcal{B}^{\text{obs}}_{l_1 l_2 l_3}$ is a subtle but important one. While in the definition of $B_{l_1 l_2 l_3}$ one  should  perform  an average over an ensemble of universes [as is explicitly stated in the definition \eqref{avg}], the object $\mathcal{B}^{\text{obs}}_{l_1 l_2 l_3}$   involves no averages over idealized  ensembles whatsoever.\footnote{One  should  not confuse  the  fact that when  obtaining the  specific  values of $ a_{lm}$, which  result from observations,   one  needs to perform an integral over   the CMB sky [as  indicated in  Eq. \eqref{alm0}],  with taking averages  over  ensembles of universes  as considered   above.} The only average that is
 being performed in $\mathcal{B}^{\text{obs}}_{l_1 l_2 l_3}$, is an average over 
orientations (i.e. a sum over $m_i$ with a weight given by the Wigner 3-$j$ symbols).

Explicitly $\mathcal{B}^{\text{obs}}_{l_1 l_2 l_3}$ is given by substituting \eqref{almrandom} in \eqref{bispecorig} which yields

\bea\label{bispecorig2}
\mathcal{B}^{\text{obs}}_{l_1 l_2 l_3} &=& \sum_{m_i}  \left( \begin{array}{lcr}
      l_1& l_2 & l_3  \\
     m_1 & m_2 & m_3
    \end{array}
    \right) \frac{ 1 }{L^{9/2}} \sum_{\nk_1, \nk_2, \nk_3} \frac{g(z_{k_1}) g(z_{k_2}) g(z_{k_3}) j_{l_1}(k_1 R_D) j_{l_2}(k_2 R_D) j_{l_3}(k_3 R_D) }{(k_1 k_2 k_3 )^{3/2}} \nonumber \\
&\times& Y_{l_1 m_1}^\star (\hat{k}_1) Y_{l_2 m_2}^\star (\hat{k}_2) Y_{l_3 m_3}^\star (\hat{k}_3) T(k_1) T(k_2) T(k_3)  X_{\nk_1} X_{\nk_2} X_{\nk_3}.
\eea
As is clear from Eq. \eqref{bispecorig2}, the collapse bispectrum is in effect a sum  of  random complex  numbers (i.e. a sum where each term is characterized by the product  $ X_{\nk_1} X_{\nk_2} X_{\nk_3}$, which is itself a complex random number), leading to what can be considered effectively as a two-dimensional (i.e. a  complex plane)  random walk. As is well known, one 
cannot give a perfect estimate for the direction of the final displacement  resulting from the random walk. 

Similarly   as  $\mathcal{B}^{\text{obs}}_{l_1 l_2 l_3}$ is characterized by the  sum  of  random variables
we cannot give a specific value for its outcome.  However,  as   we  will see,    in complete analogy  of  our  analysis of   the quantities $a_{lm}$, by focusing on the most likely value of the magnitude  $| \mathcal{B}^{\text{obs}}_{l_1 l_2 l_3}|^2$  we  will obtain  a reasonable  prediction. 

To  recapitulate,  the original  situation   corresponds  to the homogeneous and isotropic  vacuum state. When a sudden change of the 
initial state takes place due to  the collapse  (one for each mode),  the   mode    becomes characterized   by  a  fixed   value of  the corresponding  random variables; the collection of all the  values  of  such  random variables associated to all the modes characterizes, therefore,  our  single  and unique Universe (which in consequence fixes $|\mathcal{B}^{\text{obs}}_{l_1 l_2 l_3}|^2$); let us denote this set by
 
\beq\label{u}
 U = \{  X_{\nk},  X_{\nk'}, \ldots \}.
\eeq
Nevertheless, given the stochastic nature of the collapse,  we can consider that  the Universe  could  have corresponded  to  different set of  values  for the  random variables characterizing the Universe in a different manner $\tilde{U} = \{ \tilde{X}_{\nk},  \tilde{X}_{\nk'}, \ldots \}$. The collection of different sets $\{ U, \tilde U, \ldots \}$  thus  describe an hypothetical ensemble of universes.  In   making an estimate,  we will   be  assuming  that our 
Universe is a typical member of this hypothetical ensemble. Furthermore, we will make the assumption that the most likely (M.L.) value of the magnitude $|\mathcal{B}^{\text{obs}}_{l_1 l_2 l_3}|^2$ in such ensemble  comes very close to the corresponding one for our own Universe, that is 

\beq
\simeq  |\mathcal{B}^{\text{obs}}_{l_1 l_2 l_3}|^2_{\textrm{Our own Univierse}} \simeq |\mathcal{B}^{\text{obs}}_{l_1 l_2 l_3}|^2_{\textrm{M.L.}} 	
\eeq

Moreover,  we  can  simplify the estimate   by   taking the ensemble average $\overline{|\mathcal{B}^{\text{obs}}_{l_1 l_2 l_3}|^2}$ (the bar indicates that we are taking the ensemble 
average) and identify it with the most likely $|\mathcal{B}^{\text{obs}}_{l_1 l_2 l_3}|^2_{\textrm{M.L.}}$. It is needless to say that these two notions are not exactly the same for arbitrary  
kinds of ensembles; as a matter of fact, the relation between the two concepts depends on the probability distribution function (PDF) of the random variables. In principle, we do not 
know the exact PDF, as we have only access to a single realization--our own Universe--,  but  a  natural way to proceed is to assume  a normal (Gaussian) distribution for the random variable $X_{\nk}$. In such case, we can relate

\beq
|\mathcal{B}^{\text{obs}}_{l_1 l_2 l_3}|^2_{\textrm{M. L.}} = \overline{|\mathcal{B}^{\text{obs}}_{l_1 l_2 l_3}|^2}.
\eeq 
which implies that

\beq
|\mathcal{B}^{\text{obs}}_{l_1 l_2 l_3}|^2_{\textrm{Our own Universe}} = |\mathcal{B}^{\text{obs}}_{l_1 l_2 l_3}|^2_{\textrm{M. L.}} = \overline{|\mathcal{B}^{\text{obs}}_{l_1 l_2 l_3}|^2}.
\eeq

In the reminder  of this section, we will focus on computing $\overline{|\mathcal{B}^{\text{obs}}_{l_1 l_2 l_3}|^2}$.

Considering a Gaussian PDF for the random variables $X_{\nk}$ implies taking a Gaussian PDF for $x^R_{\nk}$ and $x^I_{\nk}$ (i.e. the real an imaginary parts of the complex random number $X_{\nk}$). This is, the ensemble average of the products $\overline{x_{\nk}^R x_{\nk'}^R}$ and $\overline{x_{\nk}^I x_{\nk'}^I}$ is characterized by

\beq\label{promedios}
\overline{x_{\nk}^R x_{\nk'}^R} = \delta_{\nk,\nk'} + \delta_{\nk,-\nk'}, \qquad \overline{x_{\nk}^I x_{\nk'}^I} = \delta_{\nk,\nk'} - \delta_{\nk,-\nk'}.
\eeq
Note that we have taken into account that the variables $x^R_{\nk}$ and $x^I_{\nk}$ are independent. Additionally, we have considered the correlation between the modes $\nk$ and $-\nk$ in accordance with the commutation relation given by $[\hat{a}^{R}_{\nk},\hat{a}^{R \dag}_{\nk'}]$ and $[\hat{a}^{I}_{\nk},\hat{a}^{I \dag}_{\nk'}]$ [see Eq. \eqref{creanRI}]. Given the relations \eqref{promedios}, the average for the product of two random variables $X_{\nk}$ (over the imaginary ensemble of universes) yields

\beq\label{delta1}
\overline{X_{\nk} X_{\nk'}} = \overline{(x^R_{\nk} + i x^I_{\nk})(x^R_{\nk'} + i x^I_{\nk'})} = 2 \delta_{\nk,-\nk'}.
\eeq
Furthermore, it is easy to check that 

\beq\label{delta2}
\overline{X^\star_{\nk} X^\star_{\nk'}} = \overline{X_{\nk} X_{\nk'}} = 2 \delta_{\nk,-\nk'} \qquad \textrm{and} \qquad \overline{X_{\nk} X^\star_{\nk'}} = 2 \delta_{\nk,\nk'}.
\eeq

Given the previous discussion and using Eq. \eqref{bispecorig2}, we can perform the average $\overline{|\mathcal{B}_{l_1 l_2 l_3}|^2}$

\bea\label{bismodavg}
\overline{|\mathcal{B}^{\text{obs}}_{l_1 l_2 l_3}|^2} &=&  \frac{ g(z)^6 }{L^9} \sum_{m_1, \ldots, m_6}  \left( \begin{array}{lcr}
      l_1& l_2 & l_3  \\
     m_1 & m_2 & m_3
    \end{array}
    \right)  \left( \begin{array}{lcr}
      l_1& l_2 & l_3  \\
     m_4 & m_5 & m_6
    \end{array}
    \right)  \nonumber \\
    &\times& \sum_{\nk_1, \ldots, \nk_6} \frac{j_{l_1}(k_1 R_D) j_{l_2}(k_2 R_D) j_{l_3}(k_3 R_D) }{(k_1 k_2 k_3 k_4 k_5 k_6 )^{3/2}}  j_{l_1} (k_4 R_D) j_{l_2} (k_5 R_D) j_{l_3} (k_6 R_D) Y_{l_1 m_1}^\star (\hat{k}_1)  \nonumber \\
    &\times& Y_{l_2 m_2}^\star (\hat{k}_2) Y_{l_3 m_3}^\star (\hat{k}_3)  Y_{l_1 m_4} (\hat{k}_4) Y_{l_2 m_5} (\hat{k}_5) Y_{l_3 m_6} (\hat{k}_6)  T(k_1) T(k_2) T(k_3) T(k_4) T(k_5) T(k_6) \nonumber \\
    &\times& \overline{X_{\nk_1} X_{\nk_2} X_{\nk_3} X^\star_{\nk_4} X^\star_{\nk_5} X^\star_{\nk_6}}.
\eea
where in obtaining  Eq. \eqref{bismodavg} we have  assumed $z_k$ independent of $k$ (see discussion bellow Eq. \eqref{ML6}).


As the  random variables $X_{\nk}$ are  taken to be distributed according to a Gaussian PDF, we can use the following relation

\bea\label{randoms}
\overline{X_{\nk_1} X_{\nk_2} X_{\nk_3} X^\star_{\nk_4} X^\star_{\nk_5} X^\star_{\nk_6}} &=& \nonumber \\
\overline{X_{\nk_1}^{} X_{\nk_2}^{}} \cdot \overline{X_{\nk_3}^{} X_{\nk_4}^{\star}} \cdot \overline{X_{\nk_5}^{\star} X_{\nk_6}^{\star}} &+& \overline{X_{\nk_1}^{} X_{\nk_2}^{}} \cdot \overline{X_{\nk_3}^{} X_{\nk_5}^{\star}} \cdot \overline{X_{\nk_4}^{\star} X_{\nk_6}^{\star}} + \overline{X_{\nk_1}^{} X_{\nk_2}^{}} \cdot \overline{X_{\nk_3}^{} X_{\nk_6}^{\star}} \cdot \overline{X_{\nk_4}^{\star} X_{\nk_5}^{\star}} \nonumber \\
+\overline{X_{\nk_1}^{} X_{\nk_3}^{}} \cdot \overline{X_{\nk_2}^{} X_{\nk_4}^{\star}} \cdot \overline{X_{\nk_5}^{\star} X_{\nk_6}^{\star}} &+& \overline{X_{\nk_1}^{} X_{\nk_3}^{}} \cdot \overline{X_{\nk_2}^{} X_{\nk_5}^{\star}} \cdot \overline{X_{\nk_4}^{\star} X_{\nk_6}^{\star}} + \overline{X_{\nk_1}^{} X_{\nk_3}^{}} \cdot \overline{X_{\nk_2}^{} X_{\nk_6}^{\star}} \cdot \overline{X_{\nk_4}^{\star} X_{\nk_5}^{\star}} \nonumber \\
+\overline{X_{\nk_1}^{} X_{\nk_4}^{\star}} \cdot \overline{X_{\nk_2}^{} X_{\nk_3}^{}} \cdot \overline{X_{\nk_5}^{\star} X_{\nk_6}^{\star}} &+& \overline{X_{\nk_1}^{} X_{\nk_4}^{\star}} \cdot \overline{X_{\nk_2}^{} X_{\nk_5}^{\star}} \cdot \overline{X_{\nk_3}^{} X_{\nk_6}^{\star}} + \overline{X_{\nk_1}^{} X_{\nk_4}^{\star}} \cdot \overline{X_{\nk_2}^{} X_{\nk_6}^{\star}} \cdot \overline{X_{\nk_3}^{} X_{\nk_5}^{\star}} \nonumber \\
+\overline{X_{\nk_1}^{} X_{\nk_5}^{\star}} \cdot \overline{X_{\nk_2}^{} X_{\nk_3}^{}} \cdot \overline{X_{\nk_4}^{\star} X_{\nk_6}^{\star}} &+& \overline{X_{\nk_1}^{} X_{\nk_5}^{\star}} \cdot \overline{X_{\nk_2}^{} X_{\nk_4}^{\star}} \cdot \overline{X_{\nk_3}^{} X_{\nk_6}^{\star}} + \overline{X_{\nk_1}^{} X_{\nk_5}^{\star}} \cdot \overline{X_{\nk_2}^{} X_{\nk_6}^{\star}} \cdot \overline{X_{\nk_3}^{} X_{\nk_4}^{\star}} \nonumber \\
+\overline{X_{\nk_1}^{} X_{\nk_6}^{\star}} \cdot \overline{X_{\nk_2}^{} X_{\nk_3}^{}} \cdot \overline{X_{\nk_4}^{\star} X_{\nk_5}^{\star}} &+& \overline{X_{\nk_1}^{} X_{\nk_6}^{\star}} \cdot \overline{X_{\nk_2}^{} X_{\nk_4}^{\star}} \cdot \overline{X_{\nk_3}^{} X_{\nk_5}^{\star}} + \overline{X_{\nk_1}^{} X_{\nk_6}^{\star}} \cdot \overline{X_{\nk_2}^{} X_{\nk_5}^{\star}} \cdot \overline{X_{\nk_3}^{} X_{\nk_4}^{\star}}. \nonumber \\
\eea

Using Eqs. \eqref{delta2}, we find

\bea\label{deltas}
\frac{\overline{X_{\nk_1} X_{\nk_2} X_{\nk_3} X^\star_{\nk_4} X^\star_{\nk_5} X^\star_{\nk_6}}}{2^3} &=& \nonumber \\
\delta_{\nk_1,-\nk_2} \cdot \delta_{\nk_3,\nk_4} \cdot \delta_{\nk_5,-\nk_6} &+& \delta_{\nk_1,-\nk_2} \cdot \delta_{\nk_3,\nk_5} \cdot \delta_{\nk_4,-\nk_6} + \delta_{\nk_1,-\nk_2} \cdot \delta_{\nk_3,\nk_6} \cdot \delta_{\nk_4,-\nk_5} \nonumber \\ 
+\delta_{\nk_1,-\nk_3} \cdot \delta_{\nk_2,\nk_4} \cdot \delta_{\nk_5,-\nk_6} &+& \delta_{\nk_1,-\nk_3} \cdot \delta_{\nk_2,\nk_5} \cdot \delta_{\nk_4,-\nk_6} + \delta_{\nk_1,-\nk_3} \cdot \delta_{\nk_2,\nk_6} \cdot \delta_{\nk_4,-\nk_5} \nonumber \\ 
+\delta_{\nk_1,\nk_4} \cdot \delta_{\nk_2,-\nk_3} \cdot \delta_{\nk_5,-\nk_6} &+& \delta_{\nk_1,\nk_4} \cdot \delta_{\nk_2,\nk_5} \cdot \delta_{\nk_3,\nk_6} + \delta_{\nk_1,\nk_4} \cdot \delta_{\nk_2,\nk_6} \cdot \delta_{\nk_3,\nk_5} \nonumber \\ 
+\delta_{\nk_1,\nk_5} \cdot \delta_{\nk_2,-\nk_3} \cdot \delta_{\nk_4,-\nk_6} &+& \delta_{\nk_1,\nk_5} \cdot \delta_{\nk_2,\nk_4} \cdot \delta_{\nk_3,\nk_6} + \delta_{\nk_1,\nk_5} \cdot \delta_{\nk_2,\nk_6} \cdot \delta_{\nk_3,\nk_4} \nonumber \\ 
+\delta_{\nk_1,\nk_6} \cdot \delta_{\nk_2,-\nk_3} \cdot \delta_{\nk_4,-\nk_5} &+& \delta_{\nk_1,\nk_6} \cdot \delta_{\nk_2,\nk_4} \cdot \delta_{\nk_3,\nk_5} + \delta_{\nk_1,\nk_6} \cdot \delta_{\nk_2,\nk_5} \cdot \delta_{\nk_3,\nk_4}. \nonumber \\ 
\eea

The following steps are straightforward: First, substitute \eqref{deltas} in \eqref{bismodavg} which will convert the sum over $\nk_1,...,\nk_6$ into a sum over $\nk_1,\nk_2,\nk_3$. Then, take the limit $L \to  \infty$ and $k \to$ continuum in order to switch from sums to integrals over $\nk$. Finally, perform the remaining integrals (note that we have, once again, assumed $T(k)=1$, which should be a good approximation for $l \lesssim 20$ ).
The interested reader can consult the details of the calculation in Appendix \ref{calculos}. Here, we will just present the result which is

\beq\label{bispecavg}
\overline{|\mathcal{B}^{\text{obs}}_{l_1 l_2 l_3}|^2} = \frac{ g(z)^6  ( 1+  \Delta_{l_1 l_2 l_3} ) }{(2 \pi)^9 l_1(l_1 +1) l_2(l_2 +1) l_3 (l_3+1)},
\eeq
where

\bea\label{deltagrande}
 \Delta_{l_1 l_2 l_3} &\equiv& \delta_{l_1,l_2} \left[ (2l_3+1) \left(\sum_{m_1} \left( \begin{array}{lcr}
      l_1& l_1 & l_3  \\
     m_1 & -m_1 & 0
    \end{array}
    \right)  (-1)^{m_1} \right)^2 + (-1)^{l_3+2l_1} \right] \nonumber \\
    &+& \delta_{l_2,l_3} \left[ (2l_1+1) \left(\sum_{m_2} \left( \begin{array}{lcr}
      l_2& l_2 & l_1  \\
     m_2 & -m_2 & 0
    \end{array}
    \right)  (-1)^{m_2} \right)^2 + (-1)^{l_1+2l_2} \right] \nonumber \\
    &+& \delta_{l_1,l_3} \left[ (2l_2+1) \left(\sum_{m_3} \left( \begin{array}{lcr}
      l_3& l_3 & l_2  \\
     m_3 & -m_3 & 0
    \end{array}
    \right)  (-1)^{m_3} \right)^2 + (-1)^{l_2+2l_3} \right] \nonumber \\
    &+& \delta_{l_1,l_2}\delta_{l_2,l_3} \left[ 2 + 3 (1+(-1)^{3l_1} ) (2l_1+1) \left(  \sum_{m_1}  \left( \begin{array}{lcr}
      l_1& l_1 & l_1  \\
     m_1 & -m_1 & 0
    \end{array}
    \right)      (-1)^{m_1} \right)^2 \right],
\eea
with $l=1,2, \ldots$ and $-l \leq m \leq l$. Consequently, the most likely value for the magnitude of the collapse bispectrum is  $|\mathcal{B}_{l_1 l_2 l_3}|_{\textrm{M.L.}} = \left( \overline{|\mathcal{B}_{l_1 l_2 l_3}|^2} \right)^{1/2}$, i.e.

\beq\label{bispeclm}
|\mathcal{B}_{l_1 l_2 l_3}|_{\textrm{M.L.}} = \frac{1}{ \pi^{3/2}} \left( \frac{H}{10 M_P \epsilon^{1/2}} \right)^3  \left| \cos z - \frac{\sin z}{z} \right|^3 \left( \frac{ 1+  \Delta_{l_1 l_2 l_3} }{ l_1(l_1 +1) l_2(l_2 +1) l_3 (l_3+1)} \right)^{1/2},
\eeq
where we used the definition of $g(z)$ given in Eq. \eqref{gz}.

Equation \eqref{bispeclm} is the main result of this section. Given the definition of the collapse bispectrum [see. Eq. \eqref{bispecorig}], we note that $l_1,l_2,l_3$ must correspond to the sides of a triangle, otherwise $|\mathcal{B}_{l_1 l_2 l_3}|_{\textrm{M.L.}} =0$. Furthermore, if such ``triangle'' has different side lengths ($l_1 \neq l_2 \neq l_3$), then $\Delta_{l_1 l_2 l_3}$ vanishes exactly (but not $|\mathcal{B}^{\text{obs}}_{l_1 l_2 l_3}|_{\textrm{M.L.}}$). However, if $l_1,l_2,l_3$ are associated with the sides of an isosceles ($l_i = l_j \neq l_k$) or an equilateral triangle ($l_1 = l_2 = l_3$) the terms appearing in $\Delta_{l_1 l_2 l_3}$, contribute to the collapse bispectrum, which generically does not vanishes (e.g., $|\mathcal{B}^{\text{obs}}_{2 2 2}|_{\textrm{M.L.}}^2 = g(z)^6/[(2\pi)^9 6^2]$).

We must emphasize that  $|\mathcal{B}^{\text{obs}}_{l_1 l_2 l_3}|$ is not exactly the magnitude of the traditional theoretical angle-averaged-bispectrum $|B_{l_1 l_2 l_3}|$ as the latter would correspond to take the magnitude of the object defined in Eq. \eqref{avg}. In fact, in the conventional approach, one would relate an average over an ensemble of universes with a certain quantum  three-point function and that  would  vanish  in the  absence of ``non-Gaussianities.'' Then,  the average over an ensemble of universes would be somehow related with the angle-averaged-bispectrum $B_{l_1 l_2 l_3}$, which is  a sort  of orientation average in  our Universe, connected  with    suitable averages  over   $m$'s of   the quantity $a_{l_1 m_1} a_{l_2 m_2} a_{l_2 m_3}$ measured in our own (single) Universe.

Our point is,  thus,  connected  with the  fact  that these series of identifications have an unequivocal  no clear justification (see Ref. \cite{susana2013}). On the other hand,  within the collapse model, we find a prediction for the most likely value of  $|\mathcal{B}^{\text{obs}}_{l_1 l_2 l_3}|_{\textrm{M.L.}}$ which can be related directly with the   actual observational quantity: 

\beq\label{avgbispecobs}
|B_{l_1 l_2 l_3}|_{\textrm{  Actual obs}} \equiv \left| \sum_{m_i} \left( \begin{array}{lcr}
      l_1& l_2 & l_3  \\
     m_1 & m_2 & m_3
    \end{array}
    \right) (a_{l_1 m_1} a_{l_2 m_2} a_{l_2 m_3})_{\textrm{ Actual obs}} \right|,
    \eeq
in a  direct and transparent manner. Note,  however,  that  we  make  a distinction  between the theory's  prediction  for the most likely  value of the  observational quantity $|\mathcal{B}^{\text{obs}}_{l_1 l_2 l_3}|_{\textrm{M.L.}}$  and the  actually  observed  quantity itself  $|B_{l_1 l_2 l_3}|_{\textrm{Actual obs}}$.

 In the next section, we will  extend  the discussion about the relation between the statistical aspects of the traditional bispectrum and the collapse bispectrum.

\section{Main differences between the standard and the collapse approach regarding the primordial bispectrum}\label{diferencias}

\subsection{The standard approach to the CMB bispectrum}

We begin this section by giving a rather brief review of the conventional approach to the primordial bispectrum, its amplitude and the usual arguments given to relate it  with possible non-Gaussian features in the perturbations; extended reviews can be found in Refs. \cite{bartolo2004,komatsu2003,Yadav,Liguori,Komatsu2001}.

According  to  the standard approach, if  $\Psi(\x,\eta)$ is  taken to  be  characterized  by a  Gaussian distribution all its statistical properties are codified  in the two-point correlation function. Otherwise, one needs to consider higher order correlation functions, e.g. the three-point correlation function $\overline{\Psi(\x,\eta) \Psi(\vec{y},\eta) \Psi (\vec{z},\eta)}$. The Fourier transform is commonly referred as the bispectrum, defined by

\beq\label{3ptospsi}
\overline{\Psi_{\nk_1} \Psi_{\nk_2} \Psi_{\nk_2}} \equiv (2\pi)^3 \delta (\nk_1 + \nk_2 + \nk_3) B_\Psi (k_1,k_2,k_3). 
\eeq
The Dirac delta appearing in Eq. \eqref{3ptospsi}  is said to  indicate that the ensemble average $\overline{\Psi (\vec{x},\eta) \Psi(\vec{y},\eta) \Psi(\vec{z},\eta)}$  is invariant under spatial  translations; in addition, the dependence only on the magnitudes $k$, appearing  in the function $B_\Psi (k_1,k_2,k_3)$, is  tied   to  the rotational  invariance of such ensemble. The Dirac delta  constrains  the three modes  involved; this is, the modes must satisfy $\nk_1 + \nk_2 + \nk_3 =0$, which is known as the triangle condition.  Thus,   according to  the direct relation of the Newtonian potential with  the   source of the  observed  anisotropies,  the assumption  about  the rotational and translational invariance of the ensemble  and   its  impact on $\overline{\Psi (\vec{x},\eta) \Psi(\vec{y}, \eta) \Psi(\vec{z},\eta)}$ should  be reflected  in  the invariance of the ensemble average of the temperature fluctuations $ \overline{\frac{\delta T}{T_0} (\theta_1, \varphi_1) \frac{\delta T}{T_0}             {(\
theta_2,\varphi_2)} \frac{\delta T}{T_0} (\theta_3, \varphi_3)}$.  However,  we  should   once  again  emphasize    that strictly speaking the  discussion  above  and, thus, the  average  indicated by  the overline in the latter expressions, strictly refers to an   average over an ensemble of Universes and not averages over orientations in one Universe.  That  distinction is  often ignored   in the   traditional    treatments   of the  subject.

One of the first (and most popular ways) to parameterize the traditional non-Gaussianity phenomenologically is via   the introduction of a non-linear correction to the linear Gaussian curvature perturbation
\cite{salopek1990,salopek1991},

\beq\label{psiloc}
\Psi (\x,\eta) =  \Psi_g (\x,\eta) + f_{\textrm{NL}}^\textrm{loc} [ \Psi_g^2 (\x,\eta) - \overline{\Psi_g^2 (\x,\eta)} ],
\eeq
where $\Psi_g (\x,\eta)$ denotes a linear Gaussian part of the perturbation and $\overline{\Psi_g^2 (\x,\eta)}$ is the variance of the of the Gaussian part.\footnote{Let us recall that the
variance $\overline{\Psi^2 (\x,\eta)} = \int_0^\infty dk k^2 P_\Psi(k,\eta)$ diverges logarithmically  for a power spectrum such that $P_\Psi (k,\eta) \propto k^{-3}$ unless one introduce an 
\emph{ad hoc} cutoff for $k$. For a detailed discussion of this  and other related issues see Ref. \cite{susana2013} and Appendix B of the same Ref. } 
The parameter $f_{\textrm{NL}}^{\textrm{loc}}$ is called the ``local non-linear coupling parameter" and determines the ``strength" of the primordial non-Gaussianity. This parametrization 
of non-Gaussianity is local in real space and therefore is called ``local non-Gaussianity.''  Using Eqs. \eqref{3ptospsi} and \eqref{psiloc}, the bispectrum of local non-Gaussianity may be 
derived:

\bea\label{bispeck}
B_\Psi (\nk_1,\nk_2,\nk_3) &=& 2 f_{\textrm{NL}}^\textrm{loc} [ P_\Psi (\nk_1) P_\Psi (\nk_2) + P_\Psi (\nk_2) P_\Psi (\nk_3) + P_\Psi (\nk_3) P_\Psi (\nk_1) ], \nonumber \\
&=& 2 f_{\textrm{NL}}^\textrm{loc} \frac{A^2_{\Psi}}{(k_1 k_2 k_3)^3} \left(\frac{k_1^2}{k_2k_3} + \frac{k_2^2}{k_1k_3} + \frac{k_3^2}{k_1k_2} \right),
\eea
with $P_\Psi (\nk)$ the power spectrum of the Newtonian potential  defined as $\overline{\Psi (\nk) \Psi (\nk')} \equiv (2\pi)^3 \delta (\nk + \nk') P_\Psi (k)$ and it is assumed to be of the form $P_\Psi (\nk) = A_\Psi k^{-3}$, where $A_\Psi$ denotes the amplitude of the power spectrum. 

Expressing the coefficients $a_{lm}$ in terms of $\Psi_{\nk}$ as in Eq. \eqref{alm2} and using Eq. \eqref{3ptospsi}, one can proceed to compute the CMB bispectrum, defined in Eq. \eqref{bispec0}, this is

\bea\label{fullbis}
B^{l_1 l_2 l_3}_{m_1 m_2 m_3} &=& \overline{a_{l_1 m_1} a_{l_2 m_2} a_{l_3 m_3}} \nonumber \\
&=&  \bigg( \frac{4\pi}{3} \bigg)^3 i^{l_1 + l_2 + l_3} \int \frac{d^3 k_1}{(2\pi)^3} \frac{d^3 k_2}{(2\pi)^3} \frac{d^3 k_3}{(2\pi)^3}  j_{l_1} (k_1 R_D) j_{l_2} (k_2 R_D) j_{l_3} (k_3 R_D)   \nonumber \\ 
&\times& T(k_1) T(k_2) T(k_3) \overline{\Psi_{\nk_1} \Psi_{\nk_2} \Psi_{\nk_3} } Y_{l_1 m_1} (\hat{k_1} ) Y_{l_2 m_2} (\hat{k_2} ) Y_{l_3 m_3} (\hat{k_3} ) \nonumber \\
&=& \bigg( \frac{2}{ 3 \pi} \bigg)^3  \int dk_1 dk_2 dk_3 \textrm{ } (k_1 k_2 k_3)^2  T(k_1) T(k_2) T(k_3) j_{l_1} (k_1 R_D) j_{l_2} (k_2 R_D) j_{l_3} (k_3 R_D) \nonumber \\
&\times&  B_\Psi (k_1,k_2,k_3) \int_0^\infty dx \textrm{ } x^2 j_{l_1} (k_1 x) j_{l_2} (k_2 x) j_{l_3} (k_3 x)  \int d \Omega_{\hat{x}} Y_{l_1 m_1} (\hat{x} ) Y_{l_2 m_2} (\hat{x} ) Y_{l_3 m_3} (\hat{x} ), \nonumber\\
\eea
where in the last line, one performs the integral over the angular parts of the three $k_i$ and use the exponential integral form for the delta function that appears in the bispectrum definition \eqref{3ptospsi}. The integral over the angular part of $\vec x$, in the last line, is known as the Gaunt integral, i.e.,

\begin{eqnarray}\label{gaunt}
\mathcal{G}_{l_1 l_2 l_3}^{m_1 m_2 m_3} &\equiv& \int d \Omega_{\hat{x}} Y_{l_1 m_1} (\hat{x} ) Y_{l_2 m_2} (\hat{x} ) Y_{l_3 m_3} (\hat{x} ) \nonumber \\
&=& \sqrt{\frac{(2l_1+1)(2l_2+1)(2l_3+1) }{ 4\pi } } \left( \begin{array}{lcr}
      l_1& l_2 & l_3  \\
     0 & 0 & 0
    \end{array}
    \right) \left( \begin{array}{lcr}
      l_1& l_2 & l_3  \\
     m_1 & m_2 & m_3
    \end{array}
    \right).
\end{eqnarray}
The fact that the bispectrum $B_{m_1 m_2 m_3}^{l_1 l_2 l_3}$  turns  out to be   proportional  of the Gaunt integral, $\mathcal{G}_{l_1 l_2 l_3}^{m_1 m_2 m_3}$, implies that the values of $l,m$, corresponding to the non-vanishing  components  of  the  bispectrum,  must satisfy ``the triangle conditions'' and also reflect   the fact that  in the definition of the bispectrum there  is  a Dirac delta [Eq. \eqref{3ptospsi}] which  guarantees the  rotational invariance for the ensemble average $\overline{\Psi_{\nk_1} \Psi_{\nk_2} \Psi_{\nk_3}}$. 

One can find a close form for the bispectrum by choosing to work in the Sach-Wolfe approximation, where the transfer function $T(k)=1$; thus, substituting Eq. \eqref{bispeck} into Eq. \eqref{fullbis} and evaluating  the remaining integrals one obtains (See Ref. \cite{fergusson2009}):

\bea\label{bispecloc44}
B^{l_1 l_2 l_3}_{m_1 m_2 m_3} &=& \mathcal{G}_{l_1 l_2 l_3}^{m_1 m_2 m_3} f_{\textrm{NL}}^{\textrm{loc}} \left( \frac{2 A_{\Psi}^2}{27 \pi^2} \right) \bigg( \frac{1}{l_1(l_1+1)l_2(l_2+1)} + 
\frac{1}{l_2(l_2+1)l_3(l_3+1)} \nonumber \\
&+& \frac{1}{l_1(l_1+1)l_3(l_3+1)} \bigg).
\eea
Finally, using the definition of the angle-averaged-bispectrum [Eq. \eqref{avg}], one has

\bea\label{bispecloc0}
B_{l_1 l_2 l_3} &=& \sqrt{\frac{(2l_1+1)(2l_2+1)(2l_3+1)}{4\pi}} \left( \begin{array}{lcr}
      l_1& l_2 & l_3  \\
     0 & 0 & 0
    \end{array}
    \right) f_{\textrm{NL}}^{\textrm{loc}} \left( \frac{2 A_{\Psi}^2}{27 \pi^2} \right) \nonumber \\
    &\times&  \left( \frac{1}{l_1(l_1+1)l_2(l_2+1)} + 
\frac{1}{l_2(l_2+1)l_3(l_3+1)} + \frac{1}{l_1(l_1+1)l_3(l_3+1)} \right).
\eea
When working   within  the standard scenario for slow-roll inflation (and thus a ``nearly" scale-invariant spectrum), Maldacena found that the estimate  for the amplitude of non-Gaussianities of  the local form, is  $f_{\textrm{NL}}^{\textrm{loc}} \simeq \epsilon$  \cite{maldacena2002} (this is assumed to be in the limit when $k_1 \ll k_2 \simeq k_3$, i.e. in the so-called ``squeezed''  configuration). As seen from the previous discussion, in the conventional approach, the amplitude of the primordial bispectrum and the non-Gaussian statistics for the curvature  perturbation are intrinsically related.

\subsection{Comparing the magnitude of the collapse and the traditional bispectrum}

The magnitude of $|B_{l_1 l_2 l_3}|$ is obtained from Eq. \eqref{bispecloc0}

\bea\label{bispecloc}
|B_{l_1 l_2 l_3}| &=& \sqrt{\frac{(2l_1+1)(2l_2+1)(2l_3+1)}{4\pi}} \left| \left( \begin{array}{lcr}
      l_1& l_2 & l_3  \\
     0 & 0 & 0
    \end{array}
    \right) \right| |f_{\textrm{NL}}^{\textrm{loc}}| \left( \frac{2 H^4}{27 \pi^2 M_P^4 \epsilon^2} \right) \nonumber \\
    &\times&  \left( \frac{1}{l_1(l_1+1)l_2(l_2+1)} + \frac{1}{l_2(l_2+1)l_3(l_3+1)} + \frac{1}{l_1(l_1+1)l_3(l_3+1)} \right);
\eea
where  we have used the   estimate  for the amplitude of the power spectrum for a single scalar field in the slow-roll scenario,  given by $A_\Psi \simeq H^2/(M_P^2 \epsilon)$. 

On the other hand, the magnitude of the collapse biscpectrum is given by Eq. \eqref{bispeclm}, which we will write again

\begin{equation*}\label{biscolapso}
|\mathcal{B}^{\text{obs}}_{l_1 l_2 l_3}|_{\textrm{M.L.}} = \frac{1}{ \pi^{3/2}} \left( \frac{H}{10 M_P \epsilon^{1/2}} \right)^3  \left| \cos z - \frac{\sin z}{z} \right|^3 \left( \frac{ 1+  \Delta_{l_1 l_2 l_3}}{ l_1(l_1 +1) l_2(l_2 +1) l_3 (l_3+1)} \right)^{1/2}.
\end{equation*}

Therefore we have two distinct theoretical predictions for the actual observed bispectrum $|B_{l_1 l_2 l_3}|_{\textrm{Actual obs}}$ [see Def. \eqref{avgbispecobs}]. In the standard slow-roll inflationary scenario the prediction is given by $|{B}_{l_1 l_2 l_3}|$ Eq. \eqref{bispecloc}; meanwhile, by considering the collapse hypothesis the prediction is $|\mathcal{B}^{\text{obs}}_{l_1 l_2 l_3}|_{\textrm{M.L.}}$ Eq. \eqref{bispeclm}.

The first  and most important difference is the fact that Eq.  \eqref{bispecloc}   vanishes  unless   $f_{\textrm{NL}}^{\textrm{loc}} \not=0$,  while   no such ``primordial  non-Gaussianity'' is  required for a non-vanishing value of Eq. \eqref{bispeclm}.


The second difference is that the shape of the bispectrum, i.e. its dependence on $l$, is not the same; $|\mathcal{B}^{\text{obs}}_{l_1 l_2 l_3}|_{\textrm{M.L.}}$ scales roughly as $\sim [l(l+1)]^{-3/2}$ while $|{B}_{l_1 l_2 l_3}|$ as $\sim (2l+1)^{3/2}/[l^3(l+1)^2]$. Unfortunately, the  existing  analysis  of the observational data do not focus the exact shape of the biscpectrum, but rather  on a generic  measure  of   its amplitude. However, the reported observational amplitude of the bispectrum, which in the standard picture of slow-roll inflation  corresponds to  the non-linear parameter $f_{\textrm{NL}}^{\textrm{loc}}$, depends  on the expected  shape  that emerges  from  the theoretical  estimates  of bispectrum \cite{Yadav,Liguori}. In other words,   in the standard approach,   in order to obtain an estimate for the amplitude of the bispectrum from the observational 
data, one requires a theoretical motivated shape for the bispectrum. The  observational data and theory are  strongly   interdependent. 
In  this  way,   and relying on such theoretical  considerations,  the latest results 
from \emph{Planck} mission \cite{planckng}  lead to an estimate for the amplitudes of the bispectrum for the local, equilateral, and orthogonal 
models   given by 
$f_{\textrm{NL}}^{\textrm{loc}} = 2.7 \pm 5.8, f_{\textrm{NL}}^{\textrm{equil}} = -42 \pm 75 $, and $f_{\textrm{NL}}^{\textrm{ortho}} = -25 \pm 39 $ (68\% CL statistical). On the other hand, none of the previously considered  shapes, namely the local, equilateral or orthogonal, correspond to the one given by the collapse bispectrum (in Sec \ref{comparisons}, we will deepen this discussion).

It is also interesting to note that, in the traditional framework, there is a well known result \cite{creminelli} when considering  single-field inflation named the ``consistency relation.'' This is, the value predicted for the local non-linear parameter is given by $f_{\textrm{NL}}^{\textrm{loc}} \simeq n_s-1$, with $n_s$ the ``spectral index'' of the primordial scalar fluctuations. This prediction is obtained by 
assuming a single scalar field and  no other assumptions  (within the  standard approach); in particular, it is independent of: the form of the potential, the form of the kinetic term and the initial vacuum state.   It is clear 
that for a perfect scale-invariant spectrum, that is when $n_s=1$,  the standard prediction is $f_{\textrm{NL}}^{\textrm{loc}}=0$, which  according  to standard  estimates implies the prediction $|B_{l_1 l_2 l_3}| = 0$. In contrast, within the collapse picture,   there is  no  such  close connection between  spectral index  and the collapse bispectrum. Thus, as we proved at the end of Sec. \ref{angularspectrum}, assuming a perfect scale-independent spectrum  does not imply that the  predicted  value should be  $|\mathcal{B}^{\text{obs}}_{l_1 l_2 l_3}|_{\textrm{M.L.}} = 0$; in some sense we have given a counter-example for the ``consistency relation.''

We conclude this section by noting a fundamental implication that comes from the difference between the statistical treatment in the  collapse bispectrum $|\mathcal{B}^{\text{obs}}_{l_1 l_2 l_3}|_{\textrm{M.L.}}$ and the common bispectrum $|{B}_{l_1 l_2 l_3}|$. The collapse bispectrum was obtained assuming a Gaussian PDF for the random variable $X_{\nk}$ and, as can be seen from Eq. \eqref{masterrandom}, this translates into a Gaussian distribution for the Newtonian potential $\Psi_{\nk}$. In other words, \textbf{we have taken  a Gaussian curvature perturbation and obtained a non-vanishing prediction for the observed bispectrum} $|B_{l_1 l_2 l_3}|_{\textrm{Actual obs}}$. On the other hand, assuming a  Gaussian metric perturbation--within the traditional inflationary paradigm--would have yield $|B_{l_1 l_2 l_3}|=0$, since, in this approach, one is led to conclude  that an observed non-vanishing bispectrum is an irrefutable proof of non-Gaussian statistics for the primordial perturbations. From 
the  collapse hypothesis point of view, the observed $|B_{l_1 l_2 l_3}|_{\textrm{Actual obs}}$ corresponds to just one particular realization of a random quantum process (the self-induced collapse of the wave function). Since we do not have access to other realizations (i.e. we do not have observational access to other universes) we cannot   say anything conclusive as to   whether the underlying PDF is Gaussian or not. In other words, by measuring a non-vanishing $|B_{l_1 l_2 l_3}|_{\textrm{Actual obs}}$ in our own Universe, which corresponds to a single realization of the physical process, does not necessary mean that the ensemble average $\overline{\Psi_{\nk_1} \Psi_{\nk_2} \Psi_{\nk_3}}$, is also non-vanishing and  would  consequently  proving non-Gaussianity statistics for the ensemble. 

 In fact,  just as   we do not  expect the   actual  value  of    the one  available  realization   of    $a_{lm}$   (for a fixed  value of $l$ and $m$) to  vanish  identically, even if  somehow  the  ensemble  average of such quantity  (sorting to which  we  would   have  no  access  even if an  ensemble of universes  did  exist)  would vanish,  we  should not expect the  single realization of a   bispectrum, corresponding to our   observations   of the Universe, to  vanish  identically,  even if the  average  value would  vanish.

\section{A new observational quantity obtained from the bispectrum}\label{nuevacantidad}

In this section, we will give an explicit expression for a new quantity that in principle can be measured directly. The proposed quantity was introduced first in Ref \cite{susana2013}. The main feature of this new quantity is that  there is no mixing between theory and observational data. The motivation for presenting here the explicit form of this quantity is to illustrate the 
different  kinds  of considerations   that are  natural within  different  approaches  to the  subject (the traditional one or with the collapse hypothesis). In fact, in the conventional case, by handling all averages involved (quantum averages, ensemble averages, space average, orientation averages) as  essentially the same average, 
one is  lead  to   theoretical predictions  that cannot be     clearly and directly   connected  with   the observational quantities  and,  thus,   the  discussions    about   issues  such as  non-Gaussianities tend to   become  obscure and  are    prone to   lead  to confusion. In contrast 
with the one given by considering the collapse proposal,  the   separate  issues  can be  discussed  in a clearly  independent  manner; also,  the various   kinds  of statistical  considerations  become   disentangled  and   their treatments more transparent.

We begin by recalling that, within the traditional picture, in order to obtain  a theoretical estimate  of ${B}_{l_1 l_2 l_3}$, one needs to compute $\bra \hat{\Psi}_{\nk_1} \hat{\Psi}_{\nk_2}  \hat{\Psi}_{\nk_3} \ket$, i.e. a quantum three-point function that involves quantum field operators and initial quantum states.\footnote{Much has been written in the literature on this subject; nevertheless, the goal has been always the same, that is, to calculate the three-point correlation function by using the so-called in-in formalism. Probably the most known works were produced by Maldacena \cite{maldacena2002} and Weinberg \cite{weinberg2005}, additional known works are in Refs. \cite{creminelli}  and \cite{xchen2010}. } After such calculation,  one is required to accept the identification:

\beq\label{ident}
\bra \hat{\Psi}_{\nk_1} \hat{\Psi}_{\nk_2}  \hat{\Psi}_{\nk_3} \ket = \overline{ {\Psi}_{\nk_1} {\Psi}_{\nk_2}  {\Psi}_{\nk_3} },
\eeq
with the understanding that the right hand side of Eq. \eqref{ident} is now an average over an ensemble of universes. Such identification is not clearly justified in the conventional approach; the reason is that the situation in the cosmological context is quite distinct from the laboratory setting, in which one can clearly identify the observer, the measuring device and the physical observables. Additionally,   some  of the usual lines of  the argument continues by invoking ergodic considerations  to make a further connection between ensemble 
averages and time averages; likewise, with other imprecise arguments, one is   asked to  accept  replacing the time  averages    with spatial averages and often turning, in practice, to orientation averages.

To be more precise, in the traditional approach, once the identification given by Eq. \eqref{ident} is made, one proceeds to calculate ${B}^{l_1 l_2 l_3}_{m_1 m_2 m_3} \equiv 
\overline{a_{l_1 m_1} a_{l_2 m_2} a_{l_3 m_3} }$ and that allows one to obtain $B_{l_1 l_2 l_3}$ [see Eqs. \eqref{bispec0} and \eqref{avg}].

Furthermore, if one considers the geometrical factors associated with the Gaunt integral, which comes from considering a rotationally invariant ensemble of universes (which is often 
mistakenly considered equivalently as a rotationally invariant CMB sky), then the angular bispectrum can be factorized as follows:

\beq\label{bspec}
 B^{l_1 l_2 l_3}_{m_1 m_2 m_3} = \mathcal{G}_{l_1 l_2 l_3}^{m_1 m_2 m_3}b_{l_1 l_2 l_3}.
\eeq
where $b_{l_1 l_2 l_3}$ is the so called ``reduced bispectrum.''   Note that  if we were to consider some, in principle,  arbitrary    values  of  $B^{l_1 l_2 l_3}_{m_1 m_2 m_3}$, then, in general,  it will  not  be  the case that  
   the  quantity  $b_{l_1 l_2 l_3}$  would  contain all information about the former and,  thus;  it will not be the case that  Eq.  \eqref{bspec}
 would reproduce  the original  values.

The relation between the reduced bispectrum and the angle-averaged-bispectrum is implicitly taken to be
\bea\label{avgred}
B_{l_1 l_2 l_3 } &=& \sqrt{\frac{(2l_1+1)(2l_2+1)(2l_3+1) }{ 4\pi } } \left( \begin{array}{lcr}
      l_1& l_2 & l_3  \\
     0&0 & 0
    \end{array}
    \right)  b_{l_1 l_2 l_3}.
    \eea

 As $B^{l_1 l_2 l_3}_{m_1 m_2 m_3}$ takes   explicitly into account the conditions over $l$ and $m$ codified  in  the Gaunt integral, and as,  in  the standard   approach   all  the various types of  averages  are identified,  it  is commonly stated that the reduced bispectrum  contains all 
the relevant physical information of an inflationary model. In particular, the amplitude and the shape of the bispectrum are taken as  encoded in this object.  In fact, from the previous definitions, it is straightforward to conclude that the prediction for the observational quantity $a_{l_1 m_1} a_{l_2 m_2} a_{l_3 m_3}$, in the standard picture, is simply

\beq\label{almpromedios}
\overline{a_{l_1 m_1} a_{l_2 m_2} a_{l_3 m_3}  }=  \mathcal{G}_{l_1 l_2 l_3}^{m_1 m_2 m_3} b_{l_1 l_2 l_3}.
\eeq
  which  however  is  completely inaccessible to us,   because  we have   information only about the  realization that  occurs  in our Universe.

 The  development of the collapse   approach and the fact  that  within it  we have expressions such  as that occurring in Eq. \eqref{almrandom},
 has motivated us to introduce the previous  definition   \eqref{avg}, but also the definition of a new object called the ``reduced collapse bispectrum''  as the quantity
\beq\label{Norm-avgred}
{\tilde{b}}_{l_1 l_2 l_3 } \equiv \left[ \sqrt{\frac{(2l_1+1)(2l_2+1)(2l_3+1) }{ 4\pi } } \left( \begin{array}{lcr}
      l_1& l_2 & l_3  \\
     0&0 & 0
    \end{array}
    \right) \right]^{-1} {\cal B}^{\text{obs}}_{l_1 l_2 l_3}.
\eeq
We want to remark that the quantities $\mathcal{B}^{\text{obs}}_{l_1 l_2 l_3}$ and $\tilde{b}_{l_1 l_2 l_3}$ are not exactly the same as $B_{l_1 l_2 l_3}$ and $b_{l_1 l_2 l_3}$ since in the former we are not performing any average over any ensemble of universes. The distinction is again subtle but important.

Finally, the new quantity, which was introduced in Ref. \cite{susana2013}, is called the  ``magnitude  of the bispectral  fluctuations'' and is defined  as

  \beq\label{Fluc}
       \mathcal{F}_{l_1 l_2 l_3} \equiv
        \frac{1}{( 2l_1 +1)( 2l_2 +1)( 2l_3 +1) }  \sum_{m_i} |{a_{l_1 m_1} a_{l_2 m_2} a_{l_3 m_3}} - \mathcal{G}_{l_1 l_2 l_3}^{m_1 m_2 m_3}{\tilde{b}}_{l_1 l_2 l_3 } |^2 . 
 \eeq
As mentioned in Ref. \cite{susana2013}, such quantity could be evaluated  in principle  from  a  new  type  of analysis of the exiting  data. We will focus  next in finding the theoretical predicted value for $\mathcal{F}_{l_1 l_2 l_3}$ in the traditional approach, as well as, when considering the collapse hypothesis.

In the conventional approach, the prediction for the observed $a_{l_1 m_1} a_{l_2 m_2} a_{l_3 m_3}$ would be identified with the ensemble average $\overline{a_{l_1 m_1} a_{l_2 m_2} a_{l_3 m_3}}$ which is the definition of $B^{l_1 l_2 l_3}_{m_1 m_2 m_3}$. Furthermore, by making such prediction for the observed value (i.e. $a_{l_1 m_1} a_{l_2 m_2} a_{l_3 m_3} = \overline{a_{l_1 m_1} a_{l_2 m_2} a_{l_3 m_3}}$),     then $\tilde{b}_{l_1 l_2 l_3 } = b_{l_1 l_2 l_3}$. Consequently, by Eq. \eqref{almpromedios}, the prediction for the observed 
value of $\mathcal{F}_{l_1 l_2 l_3}$, in the traditional approach, would be exactly zero. The reason behind such result, is that all the averages involved are essentially taken to be the 
same; the predicted value for the observed $a_{l_1 m_1} a_{l_2 m_2} a_{l_3 m_3}$ is the ensemble average $\overline{a_{l_1 m_1} a_{l_2 m_2} a_{l_3 m_3}}$, then this last quantity 
is made equal to $\mathcal{G}^{l_1 l_2 l_3}_{m_1 m_2 m_3}  b_{l_1 l_2 l_3} $ which is  in practice obtained  from a further 
averaging over  different orientations in our own Universe.

On the other hand, the prediction within the collapse proposal is a non-vanishing value. For a better understanding of  the  analysis  we are lead to in  our approach, we will recapitulate some of the arguments 
mentioned in Sec. \ref{colapsobispec} when dealing with the  basic  statistical aspects in our proposal. First, we are interested in the most likely value for $\mathcal{F}_{l_1 l_2 l_3}$. The most 
likely value will be the theoretical estimate  for the observed value of $\mathcal{F}_{l_1 l_2 l_3}$. Assuming  again  a Gaussian PDF for the random variables $X_{\nk}$, and thus, a 
Gaussian PDF for each $a_{lm}$ characterizing a single Universe in the imaginary ensemble, we  identify the ensemble average $\overline{\mathcal{F}_{l_1 l_2 l_3}}$ with the most likely value. 

\beq\label{Fluc2}
       \mathcal{F}_{l_1 l_2 l_3}^{\textrm{M.L.}} =
        \frac{1}{( 2l_1 +1)( 2l_2 +1)( 2l_3 +1) }  \overline{\sum_{m_i} |{a_{l_1 m_1} a_{l_2 m_2} a_{l_3 m_3}} - \mathcal{G}_{l_1 l_2 l_3}^{m_1 m_2 m_3}{\tilde{b}}_{l_1 l_2 l_3 } |^2 }. 
 \eeq
 
Once again, we will present only the final result of the calculation to not deviate the attention of the reader from the main point of this section. The interested reader can consult Appendixes  \ref{calculos} and \ref{apendiceb} for more details. Hence, the prediction for the observed value of $\mathcal{F}_{l_1 l_2 l_3}$, in our approach, is

\bea\label{Fluccol}
       \mathcal{F}_{l_1 l_2 l_3}^{\textrm{M.L.}} &=& \frac{ g(z)^6 }{(2\pi)^9  l_1(l_1+1) l_2(l_2+1) l_3(l_3+1)} \bigg( 1  - \frac{1}{( 2l_1 +1)( 2l_2 +1)( 2l_3 +1) } +\frac{2 \delta_{l_1,l_2}}{2l_1 +1}  \nonumber \\
       &+& \frac{2 \delta_{l_2,l_3}}{2l_2 +1} + \frac{2 \delta_{l_3,l_1}}{2l_3 +1}  + \frac{8 \delta_{l_1,l_2} \delta_{l_2,l_3} }{(2l_1+1)^2} - \frac{\Delta_{l_1 l_2 l_3}}{(2l_1+1) (2l_2+1) (2l_3+1) } \bigg).
\eea
with $\Delta_{l_1 l_2 l_3}$ the object defined in expression \eqref{deltagrande}. Equation \eqref{Fluccol} is the main result for this section. Such expression is valid as long as $l_1,l_2,l_3$ correspond to the sides of a triangle (otherwise $\tilde{b}_{l_1 l_2 l_3}$ would be ill defined). Appart from such triangle condition, the values of $l$ all generic. 

 Moreover, for $l_1,l_2,l_3 \gg 1$, the rest of the terms inside the parenthesis of Eq. \eqref{Fluccol} are negligible respect to the  first, thus,

\beq\label{D3l}
 l_1(l_1+1) l_2(l_2+1) l_3(l_3+1) \mathcal{F}_{l_1 l_2 l_3} \simeq \left( \frac{ H}{10 \pi^{1/2} M_P \epsilon^{1/2}} \right)^6 \bigg( \cos z   - \frac{\sin z}{z} \bigg)^6,
\eeq
where we have used the definition of $g(z)$ [Eq \eqref{gz}].

The result in Eq. \eqref{D3l} possess a similar structure as the one given by the observed angular spectrum when the Sachs-Wolfe effect is dominant, i.e.

\beq
l(l+1)C_l \simeq \textrm{constant} .
\eeq
%
Thus, having a well defined notion over which elements one is performing  the average, the predictions for the observational quantities, in this case $\mathcal{F}_{l_1 l_2 l_3}$, change substantially.

The  complete analysis of  estimators like the one introduced in this section    will be object of  future research; however,   we wished  to present it  as  an example  of the  kind of studies  that can be done in our approach, where one can clearly identify the statistical aspects of the problem at hand.

\section{Estimates  and comparisons  with observations}\label{comparisons}

   Lets  recall  that,  in  the standard  approaches, the    study of  the  bispectrum is  tied  to the  search for non-Gaussianities  and   is,    thus, based primarily  on   the  quantity
 
 \beq\label{ec1}
 \bra \hat{\Psi} (\nk_1) \hat{\Psi} (\nk_2)  \hat{\Psi} (\nk_3) \ket = \overline{\Psi(\nk_1) \Psi(\nk_2) \Psi(\nk_3)} \equiv (2\pi)^3 \delta(\nk_1 + \nk_2 + \nk_3) B (k_1,k_2,k_3) 
 \eeq

The  connection  with observations  is  made by relating $B (k_1,k_2,k_3) $ with the angle-averaged bispectrum, for which we have expression \eqref{fullbis}, this is

\bea\label{fullbis2}
B_{l_1 l_2 l_3} &=& \sum_{m_i}  \left( \begin{array}{lcr}
      l_1& l_2 & l_3  \\
     m_1 & m_2 & m_3
    \end{array}
    \right) \overline{a_{l_1 m_1} a_{l_2 m_2} a_{l_3 m_3}} \nonumber \\
 &=&  \sqrt{\frac{(2l_1+1)(2l_2+1)(2l_3+1) }{ 4\pi } } \left( \begin{array}{lcr}
      l_1& l_2 & l_3  \\
     0 & 0 & 0
    \end{array}
    \right) \nonumber \\
    &\times &\bigg( \frac{2}{ 3 \pi} \bigg)^3  \int dk_1 dk_2 dk_3 \textrm{ } (k_1 k_2 k_3)^2  T(k_1) T(k_2) T(k_3) j_{l_1} (k_1 R_D) j_{l_2} (k_2 R_D) j_{l_3} (k_3 R_D) \nonumber \\
&\times&  B_\Psi (k_1,k_2,k_3) \int_0^\infty dx \textrm{ } x^2 j_{l_1} (k_1 x) j_{l_2} (k_2 x) j_{l_3} (k_3 x). \nonumber\\
\eea
The  search for these  non-Gaussian  effects    starts  with a theoretical moldering of the form  the effects  are thought to have. We  present here  three well-known   examples  for the form   $B(k_1,k_2,k_3)$,  based on    different theoretical models,  leading to primordial ``non-Gaussianities:''
 
 \beq
 B(k_1,k_2,k_3)^{\text{loc}} = \frac{2 A_{\Psi}^2 f_{\text{NL}}^{\text{loc}}}{(k_1k_2k_3)^3} \left( \frac{k_1^2}{k_2 k_3} + \textrm{perms.} \right), 
 \eeq

  \bea
 B(k_1,k_2,k_3)^{\text{flat}} &=& 6 A_{\Psi}^2 f_{\text{NL}}^{\text{flat}} \bigg\{ \frac{1}{k_1^{4-n_s} k_2^{4-n_s}} + \frac{1}{k_2^{4-n_s} k_3^{4-n_s}} + \frac{1}{k_3^{4-n_s} k_1^{4-n_s}}  \nonumber \\
 &+& \frac{3}{(k_1k_2k_3)^{2(4-n_s)/3}} - \bigg[ \frac{1}{k_1^{(4-n_s)/3} k_2^{2(4-n_s)/3}k_3^{4-n_s} }+ (5 \text{ perms.}) \bigg] \bigg\} 
 \eea

 \beq
  B(k_1,k_2,k_3)^{\text{feat}} = \frac{6 A_{\Psi}^2 f_{\text{NL}}^{\text{feat}}}{(k_1k_2k_3)^2} \sin \left( \frac{2\pi (k_1+k_2+k_3)}{3k_c} + \phi \right).
 \eeq

 The  form  $B(k_1,k_2,k_3)^{\text{loc}}$ arises  from considering local non-linear effects  in the curvature perturbation, originated by the  slow-roll  inflation of  a  single scalar  field; $B(k_1,k_2,k_3)^{\text{flat}}$ arises  from   considering  initial  states  other than  the Bunch-Davies  vacuum and $B(k_1,k_2,k_3)^{\text{feat}}$ is   obtained by considering inflationary  potentials with a ``feature'' that might lead to strong   deviations   from  slow-roll. The quantities, $k_c$ and  $\phi$ are constant parameters of the models   that need  not concern  us  in this  discussion. 
 We  note that the dependence  on   $k_1,k_2,k_3$ differs   sharply  from case to case. The quantity $A_{\Psi}$ is  the amplitude of  the traditional  power spectrum, the observed angular spectrum $l(l+1)C_l$ fixes it to  $A_\Psi \simeq 10^{-10}$; $f_{\text{NL}}$ is    what is  known  as the    bispectrum  amplitude and it varies depending on the model. It is  clear that different forms  of  $B(k_1,k_2,k_3)$ lead to  different    $l$ dependences for the expected  observational angular  bispectrum. 
 

 Unfortunately, this means that one cannot estimate directly a generic amplitude for all different kind of models; this is because, when searching for a signal, it would be done by taking one particular model that could be very different from another.  
 

For instance, the analysis  of the data from \emph{Planck}, WMAP, etc.   do not report  the  observed   values  for   $B_{l_1 l_2 l_3}$ (or the values for $B^{l_1 l_2 l_3}_{m_1 m_2 m_3}$ for that matter),   but rather   concentrate  on the   estimates of the value   $f_{\text{NL}}$  for various  models  that have been proposed.  
In fact, in one of the recent   articles presenting the  results from the  \emph{Planck} satellite \cite{planckng},   one reads (see pg. 32 of that reference):

\begin{quotation}
``The full bispectrum for a high-resolution map cannot be evaluated explicitly because of the sheer number of operations involved, $O(l^5_{max} )$, as well as the fact that the signal will be too weak to measure in individual multipoles with any significance. Instead, we essentially use a least-squares fit to compare the
bispectrum of the observed CMB multipoles with a particular
theoretical bispectrum $b_{l_1 l_2 l_3}$. We then extract an overall ``amplitude parameter'' $f_{\text{NL}}$ for that specific template, after defining a suitable normalization convention so that we can write $b_{l_1 l_2 l_3}= f_{\text{NL}} b_{l_1 l_2 l_3}^{th}$, where  $b_{l_1 l_2 l_3}^{th}$ is defined as the value of the
theoretical bispectrum {\it ansatz} for $f_{\text{NL}} = 1$.''
\end{quotation}

  There are  various  schemes  for  estimating $f_{\text{NL}}$ but  one of the most general  methods for evaluating $f_{\text{NL}}$ (and one  employed  by the studies of  \emph{Planck}'s  data)
  relies on the following:

\beq\label{fnl}
\hat{f}_{\text{NL}} =  \frac{1}{N^2} \sum_{l_i m_i} \left( \begin{array}{lcr}
      l_1& l_2 & l_3  \\
     m_1 &m_2 & m_3
    \end{array}
    \right) \frac{ B_{l_1 l_2 l_3}^{\text{th}}}{(C_{l_1} C_{l_2} C_{l_3})_{\text{obs}}} (a_{l_1m_1}a_{l_2m_2} a_{l_3 m_3})_{\text{obs}} 
\eeq
where $\hat{f}_{\text{NL}}$ represents  the estimated    value;  $B_{l_1 l_2 l_3}^{\text{th}}$ is the input  characterizing the theoretical  model under consideration by setting  $f_{\text{NL}}=1$, $N$ is  a normalization constant  and  $(a_{l_1m_1}a_{l_2m_2} a_{l_3 m_3})_{\text{obs}}$  are  the quantities  extracted  directly  from observations  (and  so  are the  $C_l$).

For the models    previously   mentioned,  the    estimates from \emph{Planck's} data, using the  above scheme,  lead   to: 

\begin{itemize}
\item[(i)] $f_{\text{NL}}^{\text{loc}} =  2.7 \pm 5.8$ (68\% CL), thus, $|f_{\text{NL}}^{\text{loc}}| \simeq 10$. 

\item[(ii)] $f_{\text{NL}}^{\text{flat}} = 37 \pm 77$ (0.9$\sigma$). Actually, in some other models considering non-Bunch Davies vacuum states, the form of $B(k_1,k_2,k_3)^{\text{NBD}}$, as given in Eqs. (6.2) and (6.3) of Ref. \cite{xchen2007}, \emph{Planck}'s data estimate $f_{\text{NL}}^{\text{NBD}} = 155 \pm 78$ at (2.2$\sigma$) (see Table 11 of Ref. \cite{planckng} for other estimated values of $f_{\text{NL}}$ within these type of models). Therefore, $|f_{\text{NL}}^{\text{NBD}}| \simeq 10^2.$

\item[(iii)] $f_{\text{NL}}^{\text{feat}} = 434 \pm 170$ at (2.6$\sigma$) for $k_c=0.0125$ and $\phi=0$. Another estimate for the same ``feature'' model, but considering an envelope decay function of the form $\exp[-(k_1+k_2+k_3)/mk_c]$ (where $m$ is a model-dependent parameter) with a width $\Delta k = 0.015$, yields $f_{\text{NL}}^{\text{feat}} = 765 \pm 275$ at (2.8$\sigma$) for $k_c=0.01125$ and $\phi=0$ (see Tables 12 and 13 of Ref. \cite{planckng} for other estimated values of $f_{\text{NL}}$ within these type of models). Consequently, $|f_{\text{NL}}^{\text{feat}}| \simeq 10^3.$

\end{itemize}

 Items above clearly show  that   the various   functional dependencies,  assumed  for the bispectrum, lead to  dramatically  different estimates  for the bispectrum  amplitude. 

The  prediction for the bispectrum  arising  from the collapse  model is:

\begin{equation}\label{biscolapso2}
|\mathcal{B}^{\text{obs}}_{l_1 l_2 l_3}|_{\textrm{M.L.}} = \frac{1}{ \pi^{3/2}} \left( \frac{H}{10 M_P \epsilon^{1/2}} \right)^3  \left| \cos z - \frac{\sin z}{z} \right|^3 \left( \frac{ 1+  \Delta_{l_1 l_2 l_3}}{ l_1(l_1 +1) l_2(l_2 +1) l_3 (l_3+1)} \right)^{1/2}.
\end{equation}

The  most  noteworthy feature in this prediction is the  absence of any   new  and  unknown parameter characterizing   a novel  element (i.e  the model does not introduce an analogue free parameter ${f}_{\text{NL}}$ characterizing   non-Gaussianities).   In fact, we can rewrite   Eq. \eqref{biscolapso2} as
\beq\label{bispcolapso3}
|\mathcal{B}^{\text{obs}}_{l_1 l_2 l_3}|_{\textrm{M.L.}} = A^2 \tilde{f}  \left( \frac{ 1+  \Delta_{l_1 l_2 l_3}}{ l_1(l_1 +1) l_2(l_2 +1) l_3 (l_3+1)} \right)^{1/2}.
\eeq
where  $A$ was defined in Eq. \eqref{CL} and
\beq
\tilde{f} \equiv A^{-1/2}.  
\eeq
As we have shown in Sec. \ref{angularspectrum}, the observed angular spectrum fixes  $A\simeq 10^{-10}$, so the  prediction of the  collapse model is $\tilde f \simeq 10^5$. The quantity $\tilde f$ is  what  replaces, in the collapse  model, the quantity  $f_{\text{NL}}$ of the standard  analyses, and  as indicated before, there is absolutely  no adjustable parameter to be   fixed  by  observations.

  Nevertheless,  we  should,  at  this  point,  offer  a strong   warning  regarding the  way one    might  make  use of  our estimates in  comparisons  with   observations.   As  we have  argued    at the end  of Sec. \ref{colapsobispec}, in our approach,  we  can  only make  estimates  regarding  the  most likely value of the  magnitude of   the $a_{lm}$ and,  thus, of the  spectrum,   as  well  as that of  the  bispectrum;  while  in general, the   actual values  of   such quantities  would be complex.  Thus,  one cannot simply  use   expression  \eqref{biscolapso2}, together with  data  on the $a_{lm}$,  substitute  it in the   formula  \eqref{fnl} and take  that  to  represent  our model's estimate  for the quantity   $f_{\text{NL}}$. The   expected   complex  nature of the quantities    would,  without  doubt,  play  an important role  that cannot be ignored.

 As we have seen, different shapes of the bispectrum lead to rather distinct estimates for its amplitude and given that the shape of the bispectrum in our model is very different as, say, the local model, one cannot  state that $f_\text{NL}^{\text{loc}}$ should be of the same order of magnitude as $\tilde f$. Instead, one should perform a similar estimate for $\tilde f$, as the one made with the other models  using the observational data,  and compare it with the prediction $\tilde f \simeq 10^5$. This  means that, on the one  hand  the collapse  model  enjoys a  stronger predictive power,  and, on the  other,  that is completely susceptible  to  falsification by observational data.

\section{Conclusions}\label{concl}

We  have  presented  a detailed   discussion on   the  manner, in which the   study  of   essential  statistical  features  on the   CMB  spectrum,  must  be    studied  in the context of the collapse  models.  We  have  shown  the  important  differences that   arise between the analysis in this  approach  and those tied  to the standard   approaches. We  have  seen, for instance, that  the collapse models  lead to  explicit expressions  for the quantities  that are    rather directly  observable,  the  $ a_{lm}  $  [see Eq. \eqref{almrandom}], which have no  contrapart in the  usual  analyses.   Among other  advantages,  the  expression for the coefficients $a_{lm}$ [Eq. \eqref{almrandom}]   exhibits   directly the source of the randomness involved, aspects that in the standard approach  can  only be  discussed  heuristically (simply because    there,  the random  variables are not clearly identified  and  named  as in this approach).  We   have seen that this leads to  a   rather different   analysis of 
the higher   order  statistical features,  such as  the  bispectrum, which is  usually associated  with non-Gaussianities. We have  seen that    in the  collapse  scheme, one is  lead to  a  non-vanishing     expected    bispectrum,   even if  there are no primordial non-Gaussianities (i.e.    in the statistical  analysis   of  Sec. \ref{colapsobispec}  all  statistical  correlations  are  encoded  in the   two  variable correlation functions    of the   fiducial   ensemble of universes,   which are taken  as standard).  In fact,  we  have seen that the  magnitude of such  bispectrum  is predicted and it involves no  adjustable 
parameters. This makes the  model highly    predictive, and  by the  same token,   highly  susceptible  to falsification.

 One  might  be  tempted to   use  the   estimates   for the  bispectrum  amplitude obtained in the analysis of the  data  for the other models,  but as   explained in Sec. \ref{comparisons},  the fact that   the    functional  form   of the expected  bispectrum is  very different in our model   and the  models  that have been used as    for the   comparison  with observations,  invalidates  from the start  that    program.  One can   confirm this fact simply  by  noting the   sharp differences in the estimates  of $f_{\textrm{NL}}$ that are  extracted  from  the same data  for  the various  models  mentioned above.  However, the data  are    in principle available and,   thus,  testing the prediction of the collapse  models   seems    to  be   well within  reach.
 
  Nevertheless,  before even proceeding to  do this,  we  need  to  reevaluate   the predicted  bispectrum   taking into account the  effect of the transfer functions that we have   ignored  in performing the calculation leading to    the expression  \eqref{bispeclm}. That is,  in such calculation, we  replaced the  transfer functions  by   the number $1$   so  that the    integrals could be evaluated in closed   form.
  
  Thus, the  actual  comparison   of the prediction of this model  with data   will require  the    reintroduction of the  transfer functions in the evaluation.   Carrying this   out would
  involve  numerical    calculations  in order to perform the desired integrals  that will  result in the  specific   form of the     $\mathcal{B}^{\text{obs}}_{l_1 l_2 l_3}$, which will be  suitable for comparison  with 
  observations.  We  plan to  carry this  analysis in the  near future and   to  obtain the data to  contrast  with the model's   predictions. We  would view  a  reasonable match   between  
  observations  and  the model as a   strong  indication  we  are on the right track.  We  would certainly  not  expect  a complete and precise   agreement, simply due to the fact  that,  
  our model,   allows  us  to obtain only a  most likely  value  for the quantities controlled  by the random numbers associated  with the  collapse processes.  However,   the fact that   the 
  scheme  for  evaluating the  most likely value of the bispectrum is   essentially  the    same  scheme  we  used too evaluate the  spectrum, as well as the  most likely  values of the  $C_{l}$ [see  Eq. \eqref{ML6}],
  and that we found  a   very good  match  between theory and observations,  would make  it very difficult to understand  such agreement  in one case and  any  strong departure in the  second.

  Finally,  in  Sec. \ref{nuevacantidad}, we  have
 proposed   a new  quantity   for use in the study  of  higher order   statistical   features  of the CMB  data: $\mathcal{F}_{l_1 l_2 l_3}$ [see Eq. \eqref{Fluc}].
  The  estimate for this     quantity   in  any  of the standard  scenarios  would  vanish, independently of   the  functional form of the  bispectrum,  simply  because
the  standard  schemes  offer no mathematical  characterization of the   randomness, and  that   should be a key  aspect of  any  description of   something like the   distribution of the   seeds  of cosmic  structure.  In the   collapse  scheme,  we have  a  specific   expression for the   most likely value of this  quantity. Therefore,   a  comparison with the observation would  be  a nice  empirical test of the ideas  tied to our proposals.

\acknowledgments

The work of GL is supported by  postdoctoral grant from Consejo Nacional de Investigaciones Cient\'{\i}ficas y T\'{e}cnicas, Argentina.
The work  of DS was supported in
part by the CONACYT-M\'exico Grant No. 101712 and UNAM-PAPIIT Grant No. IN107412.

\appendix

\section{Details concerning the derviation of the collapse bispectrum}\label{calculos}

In this appendix we will extend the details involved in computing Eq. \eqref{bispecavg}.  We start by rewriting the coefficient $a_{lm}$, Eq. \eqref{almrandom}, as

\beq\label{apend1}
a_{lm}= i^l g(z) \sum_{\nk} F_{lm}(\nk) X_{\nk},
\eeq
where 
$g(z)$ corresponds to the definition in Eq. \eqref{gz} (we have also assumed that $\tc = z/k$, i.e. $z$ independent of $k$ and also set $T(k)=1$), and

\beq\label{apend2}
F_{lm} (\nk) \equiv \frac{ j_l (kR_D) Y_{lm}^\star (\hat k)}{k^{3/2}}.
\eeq
Therefore, the ensemble average of the squared magnitude of the collapse bispectrum is

\beq\label{apend4original}
\overline{|\mathcal{B}^{\text{obs}}_{l_1 l_2 l_3}|^2} =  \sum_{m_1,\ldots,m_6} \left( \begin{array}{lcr}
      l_1& l_2 & l_3  \\
     m_1 & m_2 & m_3
    \end{array}
    \right) \left( \begin{array}{lcr}
      l_1& l_2 & l_3  \\
     m_4 & m_5 & m_6
    \end{array}
    \right) \overline{a_{l_1 m_1} a_{l_2 m_2} a_{l_3 m_3} a_{l_1 m_4}^\star a_{l_2 m_5}^\star a_{l_3 m_6}^\star}.
    \eeq
    Let us focus on the term $\overline{a_{l_1 m_1} a_{l_2 m_2} a_{l_3 m_3} a_{l_1 m_4}^\star a_{l_2 m_5}^\star a_{l_3 m_6}^\star}$ as its value will be of use in the calculations of Appendix \ref{apendiceb}. By making use of Eqs. \eqref{apend1} and \eqref{apend2} we have

    \bea\label{apend4}
\almsix &=&\frac{g(z)^6}{L^9} \sum_{\nk_1, \ldots, \nk_6} F_{l_1 m_1}(\nk_1) F_{l_2 m_2}(\nk_2) F_{l_3 m_3}(\nk_3) \nonumber \\
&\times&  F_{l_1 m_4}^\star (\nk_4) F_{l_2 m_5}^\star (\nk_5) F_{l_3 m_6}^\star (\nk_6)  \overline{X_{\nk_1} X_{\nk_2} X_{\nk_3} X^\star_{\nk_4} X^\star_{\nk_5} X^\star_{\nk_6} }.  \nonumber \\
\eea

The next step is to  substitute \eqref{deltas} in \eqref{apend4}. The expression obtained from such substitution will contain 15 terms of triple products of Kronecker deltas

\bea\label{sumas15}
&\textrm{ }& \almsix =  \frac{2^3 g(z)^6}{L^9} \nonumber \\
&\times& \sum_{\nk_1, \ldots, \nk_6}  F_{l_1 m_1}(\nk_1) F_{l_2 m_2}(\nk_2) F_{l_3 m_3}(\nk_3) F_{l_1 m_4}^\star (\nk_4) F_{l_2 m_5}^\star (\nk_5) F_{l_3 m_6}^\star (\nk_6) \nonumber \\
&\times& ( \delta_{\nk_1,-\nk_2} \cdot \delta_{\nk_3,\nk_4} \cdot \delta_{\nk_5,-\nk_6} + \delta_{\nk_1,-\nk_2} \cdot \delta_{\nk_3,\nk_5} \cdot \delta_{\nk_4,-\nk_6} + \delta_{\nk_1,-\nk_2} \cdot \delta_{\nk_3,\nk_6} \cdot \delta_{\nk_4,-\nk_5} \nonumber \\ 
&+& \delta_{\nk_1,-\nk_3} \cdot \delta_{\nk_2,\nk_4} \cdot \delta_{\nk_5,-\nk_6} + \delta_{\nk_1,-\nk_3} \cdot \delta_{\nk_2,\nk_5} \cdot \delta_{\nk_4,-\nk_6} + \delta_{\nk_1,-\nk_3} \cdot \delta_{\nk_2,\nk_6} \cdot \delta_{\nk_4,-\nk_5} \nonumber \\ 
&+& \delta_{\nk_1,\nk_4} \cdot \delta_{\nk_2,-\nk_3} \cdot \delta_{\nk_5,-\nk_6} + \delta_{\nk_1,\nk_4} \cdot \delta_{\nk_2,\nk_5} \cdot \delta_{\nk_3,\nk_6} + \delta_{\nk_1,\nk_4} \cdot \delta_{\nk_2,\nk_6} \cdot \delta_{\nk_3,\nk_5} \nonumber \\ 
&+& \delta_{\nk_1,\nk_5} \cdot \delta_{\nk_2,-\nk_3} \cdot \delta_{\nk_4,-\nk_6} + \delta_{\nk_1,\nk_5} \cdot \delta_{\nk_2,\nk_4} \cdot \delta_{\nk_3,\nk_6} + \delta_{\nk_1,\nk_5} \cdot \delta_{\nk_2,\nk_6} \cdot \delta_{\nk_3,\nk_4} \nonumber \\ 
&+& \delta_{\nk_1,\nk_6} \cdot \delta_{\nk_2,-\nk_3} \cdot \delta_{\nk_4,-\nk_5} + \delta_{\nk_1,\nk_6} \cdot \delta_{\nk_2,\nk_4} \cdot \delta_{\nk_3,\nk_5} + \delta_{\nk_1,\nk_6} \cdot \delta_{\nk_2,\nk_5} \cdot \delta_{\nk_3,\nk_4} ). \nonumber \\ 
\eea

Let us focus on the first term 

\beq\label{primero}
\frac{2^3 g(z)^6}{L^9} \sum_{\nk_1, \ldots, \nk_6} F_{l_1 m_1}(\nk_1) F_{l_2 m_2}(\nk_2) F_{l_3 m_3}(\nk_3) F_{l_1 m_4}^\star (\nk_4) F_{l_2 m_5}^\star (\nk_5) F_{l_3 m_6}^\star (\nk_6) \delta_{\nk_1,-\nk_2} \delta_{\nk_3,\nk_4}  \delta_{\nk_5,-\nk_6}. \nonumber \\
\eeq
The Kronecker deltas in expression \eqref{primero}, will turn the sum over $\nk_1, \ldots, \nk_6$ into a sum over $\nk_1,\nk_3,\nk_5$, that is

\beq\label{apend6}
\frac{2^3 g(z)^6}{L^9} \sum_{\nk_1, \nk_3, \nk_5} F_{l_1 m_1}(\nk_1) F_{l_2 m_2}(-\nk_1) F_{l_3 m_3}(\nk_3) F_{l_1 m_4}^\star (\nk_3) F_{l_2 m_5}^\star (\nk_5) F_{l_3 m_6}^\star (-\nk_5).
\eeq
Using the parity of the spherical harmonics $Y_{lm} (-\hat{k}) = (-1)^l Y_{lm} (\hat{k})$ gives the relation $F_{l m} (-\nk) = (-1)^l F_{lm} (\nk)$; therefore expression \eqref{apend6} becomes

\beq\label{apend7}
\frac{2^3 g(z)^6 (-1)^{l_2 + l_3}}{L^9}  \sum_{\nk_1, \nk_3, \nk_5} F_{l_1 m_1}(\nk_1) F_{l_2 m_2}(\nk_1) F_{l_3 m_3}(\nk_3) F_{l_1 m_4}^\star (\nk_3) F_{l_2 m_5}^\star (\nk_5) F_{l_3 m_6}^\star (\nk_5).
\eeq
Taking the continuum limit ($L \to \infty$ and $\nk$ discrete $\to \nk$ continous) in expression \eqref{apend7} yields

\bea\label{apend8}
\frac{2^3 g(z)^6 (-1)^{l_2 + l_3}}{(2\pi)^9}  & & \left( \int d^3 k_1 F_{l_1 m_1} (\nk_1)   F_{l_2 m_2} (\nk_1)     \right) \left( \int d^3 k_3 F_{l_1 m_4}^\star (\nk_3)   F_{l_3 m_3}(\nk_3)     \right) \nonumber \\
&\times& \left( \int d^3 k_5 F_{l_2 m_5}^\star (\nk_5)   F_{l_3 m_6}^\star (\nk_5)     \right).
\eea
Focusing now on the first integral appearing in the product of expression \eqref{apend8} we have

\bea\label{apend9}
 \int d^3 k F_{l_1 m_1} (\nk)   F_{l_2 m_2} (\nk)  &=& \int d^3 k \frac{j_{l_1} (k R_D) j_{l_2} (kR_D) }{k^3} Y_{l_1 m_1} (\hat k) Y_{l_2 m_2} (\hat k) \nonumber \\
 &=& \int_0^\infty \frac{dk}{k} j_{l_1} (k R_D) j_{l_2} (kR_D) \int d\Omega  Y_{l_1 m_1} (\hat k) Y_{l_2 m_2} (\hat k) \nonumber \\
 &=& \int_0^\infty \frac{dk}{k} j_{l_1} (k R_D) j_{l_2} (kR_D) \int d\Omega  Y_{l_1 m_1} (\hat k) Y^\star_{l_2 -m_2} (\hat k) (-1)^{m_2} \nonumber \\
 &=& \int_0^\infty \frac{dk}{k} j_{l_1} (k R_D) j_{l_2} (kR_D) \delta_{l_1,l_2} \delta_{m_1, -m_2} (-1)^{m_2} \nonumber \\
 &=& (-1)^{m_2} \int_0^\infty \frac{dk}{k} j_{l_1}^2 (kR_D) \delta_{l_1,l_2} \delta_{m_1, -m_2} \nonumber \\
 &=& (-1)^{m_2} \frac{\delta_{l_1,l_2} \delta_{m_1, -m_2}}{2l_1(l_1+1)}.
\eea
In the same way, one can easily check that 

\beq\label{apend10}
\int d^3 k F^\star_{l_1 m_1} (\nk)   F^\star_{l_2 m_2} (\nk) = \int d^3 k F_{l_1 m_1} (\nk)   F_{l_2 m_2} (\nk) =  (-1)^{m_2} \frac{\delta_{l_1,l_2} \delta_{m_1, -m_2}}{2l_1(l_1+1)}
\eeq
and

\beq\label{apend10b}
\int d^3 k F_{l_1 m_1} (\nk)   F^\star_{l_2 m_2} (\nk) = \frac{\delta_{l_1,l_2} \delta_{m_1,m_2}}{2l_1(l_1+1)}.
\eeq
Therefore, using Eqs. \eqref{apend10} and \eqref{apend10b}, the expression \eqref{apend8} is

\bea\label{apend11b}
\frac{g(z)^6 (-1)^{l_2 + l_3}}{(2\pi)^9} & &  \left( \frac{(-1)^{m_2} \delta_{l_1,l_2} \delta_{m_1,-m_2}}{ l_1(l_1+1)} \right) \left( \frac{ \delta_{l_1,l_3} \delta_{m_3,m_4}}{l_3(l_3+1)} \right) \left( \frac{(-1)^{m_6} \delta_{l_2,l_3} \delta_{m_5,-m_6}}{l_2(l_2+1)} \right) \nonumber \\
&=& \frac{g(z)^6  (-1)^{m_2+m_6} \delta_{l_1,l_2} \delta_{l_2,l_3} \delta_{m_1,-m_2} \delta_{m_3,m_4} \delta_{m_5,-m_6}}{(2\pi)^9 l_1(l_1+1) l_2(l_2+1)l_3(l_3+1)} 
\eea 

One can perform similar calculations for the rest of the 14 terms contained in Eq. \eqref{sumas15}. One special term is the one involving the product $\delta_{\nk_1 \nk_4} \delta_{\nk_2 \nk_5} \delta_{\nk_3 \nk_6}$; explicitly such term is

\beq\label{special}
\frac{2^3 g(z)^6}{L^9} \sum_{\nk_1, \ldots, \nk_6} F_{l_1 m_1}(\nk_1) F_{l_2 m_2}(\nk_2) F_{l_3 m_3}(\nk_3) F_{l_1 m_4}^\star (\nk_4) F_{l_2 m_5}^\star (\nk_5) F_{l_3 m_6}^\star (\nk_6) \delta_{\nk_1 \nk_4} \delta_{\nk_2 \nk_5} \delta_{\nk_3 \nk_6}.
\eeq
In the contiuum limit, expression \eqref{special} takes the form

\bea\label{sintegral}
\frac{2^3 g(z)^6}{(2\pi)^9} & & \left(  \int d^3 k_1 F_{l_1 m_1} (\nk_1)   F^\star_{l_1 m_4} (\nk_1)  \right) \left(  \int d^3 k_2 F_{l_2 m_2} (\nk_2)   F^\star_{l_2 m_5} (\nk_2)  \right) \nonumber \\
&\times& \left(  \int d^3 k_3 F_{l_3 m_3} (\nk_3)  F^\star_{l_3 m_6} (\nk_3)  \right).
\eea
Using Eqs. \eqref{apend10b} for the integrals, expression \eqref{sintegral} is simply

\beq\label{specialf}
\frac{g(z)^6}{(2\pi)^9 l_1(l_1+1)l_2(l_2+1)l_3(l_3+1)}  \delta_{m_1,m_4} \delta_{m_2,m_5} \delta_{m_3,m_6}.
\eeq
The term given by expression \eqref{specialf} is special because there appear no Kronecker deltas that depend on $l$ (which means that if $l_1\neq l_2 \neq l_3$ this is the only term that is not vanishing). Computing the rest of the terms of Eq. \eqref{sumas15} yields

\begin{align}\label{alm6}
&\textrm{ } \almsix =   \frac{g(z)^6}{(2\pi)^9 l_1(l_1+1)l_2(l_2+1)l_3(l_3+1)} \nonumber \\
&\times [ \delta_{m_1,m_4} \delta_{m_2,m_5} \delta_{m_3,m_6}    + \delta_{l_1,l_2} ( (-1)^{m_2+m_5} \delta_{m_1,-m_2} \delta_{m_3,m_6} \delta_{m_4,-m_5} + \delta_{m_1,m_5} \delta_{m_2,m_4} \delta_{m_3,m_6} ) \nonumber \\
&+ \delta_{l_2,l_3} ( (-1)^{m_3+m_6} \delta_{m_1,m_4} \delta_{m_2,-m_3} \delta_{m_5,-m_6} + \delta_{m_1,m_4} \delta_{m_2,m_6} \delta_{m_3,m_5} ) \nonumber \\
&+ \delta_{l_3,l_1} ( (-1)^{m_3+m_6} \delta_{m_1,-m_3} \delta_{m_2,m_5} \delta_{m_4,-m_6} + \delta_{m_1,m_6} \delta_{m_2,m_5} \delta_{m_3,m_4} ) \nonumber \\
&+ \delta_{l_1,l_2} \delta_{l_2,l_3} ( (-1)^{m_2+m_6}  \delta_{m_1,-m_2}  \delta_{m_3,m_4} \delta_{m_5,-m_6} + (-1)^{m_2+m_6} \delta_{m_1,-m_2} \delta_{m_3,m_5} \delta_{m_4,-m_6}  \nonumber \\
&+ (-1)^{m_3+m_6} \delta_{m_1,-m_3} \delta_{m_2,m_4} \delta_{m_5,-m_6} + (-1)^{m_3+m_5} \delta_{m_1,-m_3} \delta_{m_2,m_6} \delta_{m_4,-m_5} \nonumber \\
&+  (-1)^{m_3+m_6} \delta_{m_1,m_5} \delta_{m_2,-m_3} \delta_{m_4,-m_6} +  \delta_{m_1,m_5} \delta_{m_2,m_6} \delta_{m_3,m_4} \nonumber \\
&+  (-1)^{m_3+m_5} \delta_{m_1,m_6} \delta_{m_2,-m_3} \delta_{m_4,-m_5} +  \delta_{m_1,m_6} \delta_{m_2,m_4} \delta_{m_3,m_5} ) ].
\end{align}

Substituting Eq. \eqref{alm6} in Eq.  \eqref{apend4original} we obtain Eq. \eqref{bispecavg}

\bea\label{specialb}
\overline{|\mathcal{B}^{\text{obs}}_{l_1 l_2 l_3}|^2} &=&  \sum_{m_1,\ldots,m_6} \left( \begin{array}{lcr}
      l_1& l_2 & l_3  \\
     m_1 & m_2 & m_3
    \end{array}
    \right) \left( \begin{array}{lcr}
      l_1& l_2 & l_3  \\
     m_4 & m_5 & m_6
    \end{array}
    \right) \overline{a_{l_1 m_1} a_{l_2 m_2} a_{l_3 m_3} a_{l_1 m_4}^\star a_{l_2 m_5}^\star a_{l_3 m_6}^\star} \nonumber \\    
    &=& \frac{ g(z)^6  ( 1+  \Delta_{l_1 l_2 l_3} ) }{(2 \pi)^9 l_1(l_1 +1) l_2(l_2 +1) l_3 (l_3+1)}
\eea
whith $\Delta_{l_1 l_2 l_3}$ as defined by Eq. \eqref{deltagrande}.

\section{General derivation of the object $\mathcal{F}_{l_1 l_2 l_3}^\textrm{M.L.}$}\label{apendiceb}

In this appendix we will extend the details involved in computing the object $\mathcal{F}_{l_1 l_2 l_3}^{\textrm{M.L.}}$ as expressed in Eq. \eqref{Fluc2} that is 

  \beq\label{apendb1}
       \mathcal{F}_{l_1 l_2 l_3}^{\textrm{M.L.}} =
        \frac{1}{( 2l_1 +1)( 2l_2 +1)( 2l_3 +1) }  \overline{\sum_{m_i} |{a_{l_1 m_1} a_{l_2 m_2} a_{l_3 m_3}} - \mathcal{G}_{l_1 l_2 l_3}^{m_1 m_2 m_3}{\tilde{b}}_{l_1 l_2 l_3 } |^2 }. 
\eeq
%
 Using the defintion of $\tilde{b}_{l_1 l_2 l_3}$ [Eq. \eqref{Norm-avgred}], $\mathcal{F}_{l_1 l_2 l_3}^{\textrm{M.L.}}$ takes the follwing form:

\bea\label{apendb4}
\mathcal{F}_{l_1 l_2 l_3}^{\textrm{M.L.}} &=&
        \frac{1}{( 2l_1 +1)( 2l_2 +1)( 2l_3 +1) } \bigg[ \sum_{m_1,m_2,m_3} \overline{|a_{l_1 m_1} a_{l_2 m_2} a_{l_3 m_3}|^2 } \nonumber \\
        &-&  \sum_{m_1,\ldots,m_6} \left( \begin{array}{lcr}
      l_1& l_2 & l_3  \\
     m_1 & m_2 & m_3
    \end{array}
    \right) \left( \begin{array}{lcr}
      l_1& l_2 & l_3  \\
     m_4 & m_5 & m_6
    \end{array}
    \right) \overline{a_{l_1 m_1} a_{l_2 m_2} a_{l_3 m_3} a_{l_1 m_4}^\star a_{l_2 m_5}^\star a_{l_3 m_6}^\star} \bigg].
\eea

The first term in the square brackets of Eq. \eqref{apendb4} can be obtained from Eq. \eqref{alm6} by taking $m_4=m_1, m_5=m_2, m_3=m_6$; this is,

\bea\label{apendb5}
\sum_{m_1,m_2,m_3} & & \overline{|a_{l_1 m_1} a_{l_2 m_2} a_{l_3 m_3}|^2 } =  \frac{g(z)^6}{(2\pi)^9 l_1(l_1+1)l_2(l_2+1)l_3(l_3+1)} \nonumber \\
&\times & \sum_{m_1 m_2 m_3} \big[ 1 + \delta_{l_1,l_2} (\delta_{m_1,m_2} + \delta_{m_1,-m_2}) +  \delta_{l_2,l_3} (\delta_{m_2,m_3}+ \delta_{m_2,-m_3}) \nonumber \\
&+&  \delta_{l_1,l_3} (\delta_{m_1,m_3} + \delta_{m_1,-m_3}) +2 \delta_{l_1,l_2} \delta_{l_2,l_3} [ \delta_{m_1,m_2} ( \delta_{m_2,m_3} + \delta_{m_2,-m_3} )\nonumber \\
&+&  \delta_{m_1,-m_2} ( \delta_{m_2,m_3} + \delta_{m_2,-m_3} ) \big] \nonumber \\
&=& \frac{g(z)^6}{(2\pi)^9 l_1(l_1+1)l_2(l_2+1)l_3(l_3+1)}  \bigg[ (2l_1+1) (2l_2+1)(2l_3+1)  \nonumber \\
&+& 2\delta_{l_1,l_2} (2l_3+1)(2l_1+1) + 2\delta_{l_2,l_3} (2l_1+1)(2l_1+2)  + 2\delta_{l_3,l_1} (2l_2+1)(2l_3+1)  \nonumber \\
&+& 8 \delta_{l_1,l_2} \delta_{l_2,l_3} (2l_1+1) \bigg].
\eea

The second term in the square brakets of Eq. \eqref{apendb4} is exactly what we found in Eq. \eqref{specialb}. Therefore, we obtian

\bea\label{apendb6}
       \mathcal{F}_{l_1 l_2 l_3}^{\textrm{M.L.}} &=& \frac{ g(z)^6 }{(2\pi)^9  l_1(l_1+1) l_2(l_2+1) l_3(l_3+1)} \bigg( 1  - \frac{1}{( 2l_1 +1)( 2l_2 +1)( 2l_3 +1) } +\frac{2 \delta_{l_1,l_2}}{2l_1 +1}  \nonumber \\
       &+&  \frac{2 \delta_{l_2,l_3}}{2l_2 +1} + \frac{2 \delta_{l_3,l_1}}{2l_3 +1} + \frac{8 \delta_{l_1,l_2} \delta_{l_2,l_3} }{(2l_1+1)^2} - \frac{\Delta_{l_1 l_2 l_3}}{(2l_1+1) (2l_2+1) (2l_3+1) } \bigg),
\eea
which is Eq. \eqref{Fluccol}.


\end{document}